\documentclass[10pt]{IEEEtran}

\usepackage{comment}

\usepackage{times}
\usepackage{bm}
\usepackage{color}
\usepackage{theorem}
\usepackage{amsmath}
\usepackage{amssymb}
\usepackage{subfigure}
\usepackage{graphicx}
\usepackage{hyperref}
\usepackage[font=small,format=plain,up,up]{caption}
\usepackage{cite}
\usepackage[normalem]{ulem}
\usepackage{tikz,pgfplots}

\usepackage{needspace}

\input{mysymbol.sty}

\newtheorem{mytheorem}{\bf Theorem}

\newtheorem{problem}{\bf Problem}
\newtheorem{myproposition}{\bf Proposition}

\def\b{\mathbf{b}}   \def\B{\mathbf{B}}
   
\def\d{\mathbf{d}}

\def\h{\mathbf{h}}   
\def\I{\mathbf{I}}

\def\p{\mathbf{p}}   \def\P{\mathbf{P}}
   
\def\r{\mathbf{r}}    
   \def\S{\mathbf{S}}
     
   \def\U{\mathbf{U}}
\def\V{\mathbf{V}}
\def\w{\mathbf{w}} 
\def\x{\mathbf{x}}   \def\X{\mathbf{X}}
\def\y{\mathbf{y}}   \def\Y{\mathbf{Y}}
\def\z{\mathbf{z}}   \def\Z{\mathbf{Z}}

\def\E{\mathcal{E}}
\def\G{\mathcal{G}}
\def\N{\mathcal{N}}

\def\R{\mathbb{R}}

\title{Estimating Network Processes via Blind Identification of Multiple Graph Filters}
\author{Yu Zhu, Fernando J. Iglesias, Antonio G. Marques, and Santiago Segarra\thanks{Y. Zhu and S. Segarra are with the Department of Electrical and Computer Engineering, Rice University. F. J. Iglesias and A. G. Marques are with the Department of Signal Theory and Communications, King Juan Carlos University. Emails: {yz126@rice.edu, fj.iglesias@alumnos.urjc.es, antonio.garcia.marques@urjc.es, segarra@rice.edu}. This work was supported in part by the Spanish grants MINECO Klinilycs TEC2016-75361-R and Instituto de Salud Carlos III DTS17/00158. Preliminary results were published in a conference version of this paper~\cite{zhu_2019_estimation}.}}
		
\begin{document}
\maketitle
\begin{abstract}

	This paper studies the problem of jointly estimating multiple network processes driven by a common unknown input, thus effectively generalizing the classical blind multi-channel identification problem to graphs.
	%
	More precisely, we model network processes as graph filters and consider the observation of multiple graph signals corresponding to outputs of different filters defined on a common graph and driven by the same input.
	%
	Assuming that the underlying graph is known and the input is unknown, our goal is to recover the specifications of the network processes, namely the coefficients of the graph filters, only relying on the observation of the outputs.
	Being generated by the same input, these outputs are intimately related and we leverage this relationship for our estimation purposes.
	Two settings are considered, one where the orders of the filters are known and another one where they are not known.
	For the former setting, we present a least-squares approach and provide conditions for recovery.
	For the latter scenario, we propose a sparse recovery algorithm with theoretical performance guarantees.
	Numerical experiments illustrate the effectiveness of the proposed algorithms, the influence of different parameter settings on the estimation performance, and the validity of our theoretical claims.
\end{abstract}

\section{Introduction}\label{S:intro}

Since networks encode pairwise relationships between a set of agents, they can be used to model a wide range of real-world systems, with relevant examples including  social~\cite{garas_2010_worldwide, liben_2007_link}, technological~\cite{kleinberg_1999_authorative, balthrop_2004_technological}, and biological~\cite{bu_2003_topological, medaglia_2017_brain} networks.
Currently, an increasing amount of data associated with the agents (nodes) that form these networks is being collected, thus creating the pressing need to better understand graph-structured data.
The modeling and analysis of such structured data have attracted the attention from different disciplines that include statistics, machine learning, and signal processing. Within the latter, graph signal processing (GSP), a research area that seeks to generalize concepts and tools in classical digital signal processing (DSP) to data (signals) defined on a graph, has been gaining traction \cite{sandryhaila_2013_discrete,shuman_2013_emerging,ortega_2018_graph}.
Examples of relevant problems that have been recently addressed using GSP tools include graph signal sampling~\cite{marques_2016_sampling, chen_2015_sampling, anis_2016_sampling, chamon_2018_sampling, romero_2017_reconstruction}, graph-based wavelet and Fourier transforms~\cite{hammond_2011_wavelet, shafipour_2017_fourier}, and topology identification~\cite{dong_2016_laplacian, kalofolias_2016_smooth, segarra_2017_topo, shen_2017_kernel, mateos_2019_connecting}, to name a few. 

In GSP, the data collected from the nodes of the network is referred to as the graph signal, and the network topology is captured by the so-termed graph shift operator (GSO) which is the graph counterpart of the time shift in DSP \cite{sandryhaila_2013_discrete}. 
The \emph{graph filter}, a linear transformation between graph signals, generalizes the classical notion of a linear time-invariant system~\cite{sandryhaila_2013_discrete}. 
Graph filters can represent local interactions among nodes and, hence, offer the ability to model network diffusion processes, which allows us to apply signal processing tools to the study of network science problems.
In the past years, substantial effort has been devoted to the development and understanding of graph filters, including their optimal design under different metrics~\cite{segarra_2017_filters, isufi_2017_filters, teke_2017_filters}, the synthesis of graph filter banks~\cite{narang_2012_filter, sakiyama_2014_oversampled, tay_2015_design}, and the consideration of non-linear filters~\cite{segarra_2017_design, xiao_2018_nonlinear, segarra_2016_center}.

By modeling network diffusion processes as graph filters, we recast the estimation of network processes as a problem of blind identification of graph filters.
In particular, we seek to (blindly) estimate the filter coefficients by observing the filter output with no knowledge of the input.
Relevant examples of network processes include the spread of an epidemic disease in a population and the evolution of opinions in a social network. 
Correspondingly, we might be concerned with questions such as how fast the disease spreads or how interactions between people affect the formation of their opinions. 
Ultimately, answering these questions can provide insights on how to design strategies to constrain the transmission of the disease, or to avoid the proliferation of violent ideas. 
Making the connection with the graph-filter formalism, in the opinion formation example, the filter's input, output, and coefficients correspond to people's original opinions, the opinions after interaction with their peers, and how open each person is to be influenced by their peers when updating his own opinion, respectively.

The problem of blind identification of graph filters has been studied in the past \cite{segarra_2016_blind, ramirez_2017_graph, mateos_2018_blind, iglesias_demixing_2018}.
The schemes in \cite{segarra_2016_blind} and \cite{ramirez_2017_graph} seek to recover the coefficients of a single graph filter from a single output under the assumption that the unknown input is sparse. 
In \cite{mateos_2018_blind}, the authors consider a single graph filter excited by multiple sparse inputs, each of them giving rise to a different output. 
Moreover, \cite{iglesias_demixing_2018} addresses the case where a single observation formed by the sum of multiple outputs is available, and it is assumed that these outputs are generated by different sparse inputs diffused through different graph filters. 

However, in many scenarios, we have access to the outputs of \emph{multiple} related network processes excited by a common input, such as different patterns of brain activity when an individual is presented with the same visual stimulation.
Alternatively, we can think of a single network diffusion process that is sensed at different points in time as corresponding to multiple network diffusions  of different durations driven by the same input. 
For both cases, the common input motivates a \emph{joint} estimation formulation, which is the focus of this paper. 
Formally, this paper aims at recovering the coefficients of \textit{multiple graph filters} defined on one GSO and driven by a \textit{common unknown input} from the observation of their outputs. 
Moreover, the sparse input assumption is not required here, further departing from previous works.

The problem of blind identification of multiple graph filters is a generalization of the classical blind multi-channel identification problem in DSP, which is well studied and solutions have been provided under different assumptions.
Specifically, some approaches leverage second-order~\cite{tong_1991_2order,tong_1994_blind} or higher-order~\cite{Benveniste1980,Ding1991,Tugnait1987,shalvi_1990_new,Hatzinakos1989,Petropulu1993, giannakis_1989_identification, giannakis_1989_cumulant} statistical knowledge of the input for the recovery.
However, these statistics-based algorithms might suffer from model mismatch when the number of observations is limited.
To solve this problem, subspace-based algorithms~\cite{liu_1993_deterministic,liu1994,xu_1995_least,Moulines1995,Slock1994,Baccala1994} have been proposed, whose success reveals that the intrinsic single-input multiple-output structure is essential to blind identification~\cite{liu_1996_recent}.
Our work follows this direction and requires no knowledge about the input.

\emph{Contributions and paper organization:} This paper studies the problem of estimating network processes from the observation of their outputs.
By leveraging GSP tools and modeling network processes as graph filters, the problem is recast as one of blind identification of multiple graph filters. 
Moreover, we consider two settings: one where the filter orders are known and another one where they are not. 
For the setting of known filter orders, a least-squares approach is presented along with sufficient, necessary, as well as necessary and sufficient conditions for recovery. 
For the setting of unknown filter orders, we overshoot the filter orders and propose a sparse recovery algorithm; performance guarantees for identifiability and robust recovery are also provided. 

%
%
%
%
%

The rest of the paper is organized as follows. 
In Section~\ref{S:preliminaries}, we first present basic concepts in GSP, then briefly introduce the classical problem of blind multi-channel identification, and finally state our problem formulation.
In Sections \ref{S:multi_processes} and \ref{S:single_process}, the two problems of estimation of multiple network processes and a single network process are respectively discussed. 
Within each of these two sections, we tackle the two scenarios where the filter orders are respectively known and unknown.
Experimental results are presented in Section \ref{S:num_exp} and closing remarks are included in Section \ref{S:conclusions}.

\emph{Notation:} The entries of a matrix $\mathbf{X}$ and a vector $\mathbf{x}$ are denoted as $[\mathbf{X}]_{ij}$ and $x_i$, respectively.
%
%
Operations $(\cdot)^{\top}$ and $(\cdot)^{\dag}$ represent matrix transpose and pseudo-inverse, respectively.
$\mathrm{diag}(\mathbf{x})$ is a diagonal matrix whose $i$th diagonal entry is $x_i$.
$\mathbf{0}$ and $\mathbf{0}_{m\times n}$ refer to the vector and the $m\times n$ matrix whose entries are all zero.
$\|\cdot\|_p$ denotes the $\ell_p$ norm of the argument vector or matrix.
$\mathbf{x}_{\mathcal{I}}$ denotes the vector formed by the entries of $\mathbf{x}$ indexed by $\mathcal{I}$.
$\mathbf{X}_{\mathcal{I}}$ denotes the submatrix formed by the columns of $\mathbf{X}$ indexed by $\mathcal{I}$.
We use $\mathbf{X}_{\mathcal{I}}^{\top}$ to represent the transpose of $\mathbf{X}_{\mathcal{I}}$ rather than a submatrix of $\mathbf{X}^{\top}$.

\section{Preliminaries and Problem Formulation}\label{S:preliminaries}

\subsection{Fundamentals of Graph Signal Processing}\label{SS:GSP}

A directed graph $\G$ consists of a node set $\N$ of cardinality $N$ and an edge set $\E$ such that the ordered pair $(i,j)$ belongs to $\E$ if there exists an edge from node $i$ to node $j$.
The incoming neighborhood of node $i$ is defined as the set $\N_i := \{j|(j,i)\in\E\}$.
A graph signal defined on $\G$ can be represented as a vector $\x=[x_1,\cdots,x_N]^{\top}\in\R^N$, where $x_i$ denotes the signal value associated with node $i$.
The graph structure is captured by the GSO $\S\in\R^{N\times N}$ \cite{sandryhaila_2013_discrete}, whose entry $[\S]_{ji}$ can be non-zero only if $(i,j)\in\E$ or $i=j$.
Notice that $\S$ reflects the local connectivity of $\G$. 
Specifically, if $\y=\S\x$, node $i$ can compute $y_i$ as a linear combination of $x_j$ where $j\in\N_i$.
Typical choices for $\S$ are the adjacency matrix \cite{sandryhaila_2013_discrete} and the graph Laplacian~\cite{shuman_2013_emerging}. 
We assume that $\S$ is diagonalizable, so that $\S=\V\mathbf{\Lambda}\V^{-1}$, where $\V$ collects the eigenvectors of $\S$ as columns and $\mathbf{\Lambda}=\diag(\pmb{\lambda})$ collects the eigenvalues $\pmb{\lambda}=[\lambda_1,\cdots,\lambda_N]^{\top}$. 
Following the DSP terminology, the distinct eigenvalues and the eigenvectors of $\S$ are called the graph frequencies and frequency components respectively \cite{sandryhaila_2013_discrete}.
Linear and shift-invariant graph filters can be expressed as polynomials of $\S$ \cite{sandryhaila_2013_discrete}
	\begin{equation}\label{E:graph_filter}
		\mathbf{H} := \sum_{l=0}^{L-1} h_l \S^{l}.
	\end{equation}
For a given input $\x$, the output of the graph filter is given by $\y=\mathbf{H}\x$. 
Define the vector $\h := [h_0,\cdots,h_{L-1}]^{\top}$ to collect the filter coefficients, where $L$ is called the order of the filter (i.e., the size of $\h$) and $L-1$ the degree of the filter (i.e., the degree of the polynomial $\mathbf{H}$).
Notice that, any graph filter defined as in \eqref{E:graph_filter} has a unique equivalent filter on the same graph whose order is at most $N_{\S}$ where $N_{\S}$ denotes the degree of the minimal polynomial of $\S$ \cite[Thm. 3]{sandryhaila_2013_discrete}.
For diagonalizable $\S$, $N_{\S}$ equals the number of distinct eigenvalues of $\S$.
We assume that $L\leq N_{\S}$ throughout the paper.

Graph signals and filters can also be represented in the frequency domain. 
Define $\U := \V^{-1}$ and $\mathbf{\Psi}$ as a Vandermonde matrix of dimension $N\times L$ where $[\mathbf{\Psi}]_{ij} := \lambda_i^{j-1}$, then the frequency representations of a signal $\x$ and a filter $\h$ are defined as $\tilde{\x} := \U\x$ and $\tilde{\h} := \mathbf{\Psi}\h$, respectively \cite{sandryhaila_2013_discrete}. 
Correspondingly, graph filters can be rewritten as $\mathbf{H}=\U^{-1}(\sum_{l=0}^{L-1}h_l\mathbf{\Lambda}^l)\U=\U^{-1}\diag(\mathbf{\Psi}\h)\U$, and the frequency representation of the output is given by 
	\begin{equation}\label{E:output_f}
		\tilde{\y}=\U\y=\U\mathbf{H}\x=\diag(\mathbf{\Psi}\h)\U\x=\diag(\tilde{\h})\tilde{\x}=\tilde{\h}\circ\tilde{\x},
	\end{equation}
with \eqref{E:output_f} being the graph counterpart of the classical convolution theorem.

\subsection{Classical Blind Multi-Channel Identification}\label{SS:classical_BCI}

Consider the discrete multi-channel system~\cite{xu_1995_least} formed by one deterministic input signal $x(\cdot)$ of length $N$, $M$ channels represented by finite impulse responses $\{h_m(\cdot)\}_{m=1}^M$ of maximum degree $L$, and $M$ outputs $\{y_m(\cdot)\}_{m=1}^M$. 
In the noise-free case, they are related as
	\begin{equation}\label{E:classical_multi_channel}
		y_m(t) = \sum_{l=0}^L h_m(l)x(t-l),\quad m=1,\cdots,M.
	\end{equation}
The blind identification problem can be stated as follows: Given the outputs $\{y_m(t), m=1,\cdots,M; t=L,\cdots,N\}$, determine the channel coefficients $\{h_m(\cdot)\}_{m=1}^M$ without knowing the common input $x(\cdot)$. 
Notice that (\ref{E:classical_multi_channel}) can be compactly written as $y_m(t)=h_m(t)*x(t)$, where $*$ denotes the classical convolution operation. Thus, for any pair of two noise-free outputs $y_m(t)$ and $y_n(t)$, we have
	\begin{equation}\label{E:classical_equality}
		h_n(t) * y_m(t) = h_n(t) * h_m(t) * x(t) = h_m(t) * y_n(t),
	\end{equation} 
which can also be written in matrix form
	\begin{equation}\label{E:classical_equality_matrix}
		\left[\begin{matrix}\B_m & -\B_n\end{matrix}\right]\left[\begin{matrix}\h_n \\ \h_m \end{matrix}\right] = \mathbf{0},
	\end{equation} 
where $\h_m := [h_m(L),\cdots,h_m(0)]^{\top}$, and $\B_m$ is a Hankel matrix of dimension $(N-2L+1)\times(L+1)$ whose $(i,j)$ entry is $[\B_m]_{ij}:=y_m(i+j+L-2)$.
To leverage all the possible cross relations simultaneously, we apply the Data Selection Transform (DST) $\mathcal{D}$~\cite{liu_1996_recent} to $\{\B_m\}_{m=1}^M$.
In particular, define $\mathcal{D}(\{\B_{m'}\}_{m'=1}^2):=[\B_2,-\B_1]$, then $\mathcal{D}(\{\B_{m'}\}_{m'=1}^m)$ for $2<m\leq M$ are recursively given as
	\begin{align}\label{E:DST}
		\mathcal{D}(\{\B_{m'}\}_{m'=1}^m) &:= \left[\begin{array}{ccc}
				\mathcal{D}(\{\B_{m'}\}_{m'=1}^{m-1}) &\vline& \mathbf{0}\\
		\hline
		\begin{array}{ccc}
		\B_m & &  \\
		& \ddots & \\
		& & \B_m
		\end{array} &\vline&
		\begin{array}{c}
		-\B_1   \\
		\vdots  \\
		-\B_{m-1}
		\end{array}
		\end{array}\right].
	\end{align}
Consequently, all the possible cross relations of the form in (\ref{E:classical_equality_matrix}) can be condensed in   
\begin{equation}\label{E:classical_equality_final}
		\B\h = \mathbf{0},
	\end{equation}
	where $\B := \mathcal{D}(\{\B_{m}\}_{m=1}^M)$ and $\h := [\h_1^{\top},\cdots,\h_M^{\top}]^{\top}$.
If $\B$ has a one-dimensional null space, the channel coefficients $\h$ can be recovered uniquely (up to a scalar multiple) from \eqref{E:classical_equality_final}.

\subsection{Problem Formulation}\label{SS:ps}

Using the previously introduced notions, we can formalize the problem to be addressed.
Consider $M$ graph filters based on one common GSO $\S$.
For the $m$th graph filter, let $L_m$ denote its order and $\h^{(m)}:=[h^{(m)}_0,h^{(m)}_1,\cdots,h^{(m)}_{L_m-1}]^{\top}$ collect its coefficients. 
The outputs generated by these filters when excited by a common input $\x$ are given by
\begin{equation}\label{E:output_multichanel_graphfilters}
\y^{(m)}:=\mathbf{H}^{(m)}\x, \;\text{with}\;\mathbf{H}^{(m)}:=\sum_{l=0}^{L_m-1}h^{(m)}_{l}\S^{l},\;\;m=1,..., M.
\end{equation}

\begin{problem}\label{P:problem_statement}
(Blind Identification of Multiple Graph Filters)\\ 
Given $\S$ and the outputs $\{\y^{(m)}\}_{m=1}^M$ adhering to the model in \eqref{E:output_multichanel_graphfilters}, identify the filter coefficients $\{\h^{(m)}\}_{m=1}^M$ with no knowledge of the common input $\x$.
\end{problem}

We define the solution to Problem~\ref{P:problem_statement} to be \emph{identifiable} if it can be determined up to a scalar multiple, since the observed outputs keep unchanged if we multiply all the filter coefficients $\{\h^{(m)}\}_{m=1}^M$ by a scalar $\alpha$ and multiply $\x$ by $1/\alpha$ simultaneously.
To see the practical relevance of this problem, notice that graph filters can be adopted to model linear diffusion dynamics which depend on the network topology \cite{segarra_2017_filters, isufi_2017_filters, segarra_2016_blind}, with the filter coefficients and the filter orders corresponding to the diffusion rates and durations, respectively. 
Potential applications range from social networks where a rumor is spread across the network via local opinion exchanges, to brain networks where an epileptic seizure emanating from few regions is later diffused across the entire brain.
The outputs $\{\y^{(m)}\}_{m=1}^M$ might be sampled from multiple processes run on the same network or a single network process at different points in time, and we will study these two cases in Sections~\ref{S:multi_processes} and \ref{S:single_process}, respectively.
For the former case, we make no assumption on the filter coefficients, while for the latter case, the filter coefficients are dependent. 
This implies that the estimation problem of a single network process can also be seen as a special case of the more general problem of joint estimation of multiple network processes.
For each case, we further consider two settings depending on whether the filter orders $\{L_m\}_{m=1}^M$ are known or not. 
When the filter orders are given, a least-squares approach is advocated, whereas a sparse recovery algorithm is proposed for the more challenging setting of unknown filter orders. 
Identifiability conditions and theoretical guarantees are also discussed. 

\section{Joint Estimation of Multiple Network Processes}\label{S:multi_processes}

\subsection{Known Filter Orders}\label{SS:multi_known}

The knowledge of the filter orders $\{L_m\}_{m=1}^M$ allows us to define a series of Vandermonde matrices $\{\mathbf{\Psi}^{(m)}\}_{m=1}^M$, where $\mathbf{\Psi}^{(m)}$ is of dimension $N\times L_m$ and $[\mathbf{\Psi}^{(m)}]_{ij}:=\lambda_i^{j-1}$.
From this point onward, we reserve $\mathbf{\Psi}$ to represent a block diagonal matrix whose $M$ main-diagonal blocks are respectively given by $\{\mathbf{\Psi}^{(m)}\}_{m=1}^M$. 
Moreover, we define a series of matrices $\{\tilde{\Y}^{(m)}\}_{m=1}^M$, where $\tilde{\Y}^{(m)} := \diag(\tilde{\y}^{(m)})$ and $\tilde{\y}^{(m)}$ is the frequency representation of the $m$th output $\y^{(m)}$.
Applying the DST defined in (\ref{E:DST}) to $\{\tilde{\Y}^{(m)}\}_{m=1}^M$, we further define $\tilde{\Y} := \mathcal{D}(\{\tilde{\Y}^{(m)}\}_{m=1}^M)$.

\begin{myproposition}\label{Prop:multi}
Defining $\h:=[\h^{(1)\top},\h^{(2)\top},\cdots,\h^{(M)\top}]^{\top}$ obtained by vertically concatenating the coefficients of the $M$ unknown graph filters, the following expression holds
	\begin{equation}\label{E:multi_basic}
		\tilde{\Y}\mathbf{\Psi}\h=\mathbf{0}.
	\end{equation}
\end{myproposition}

\begin{myproof}
	From (\ref{E:output_f}) it follows that $\tilde{\y}^{(m)}=\tilde{\h}^{(m)}\circ\tilde{\x}$ for each filter $m$, where $\tilde{\h}^{(m)}$ is the frequency response of the $m$th filter.
In a spirit similar to (\ref{E:classical_equality}), we have that
	\begin{equation}\label{E:multi_pair}
		\tilde{\h}^{(n)} \circ \tilde{\y}^{(m)} = \tilde{\h}^{(n)} \circ \tilde{\h}^{(m)} \circ \tilde{\x} = \tilde{\h}^{(m)} \circ \tilde{\h}^{(n)} \circ \tilde{\x} = \tilde{\h}^{(m)} \circ \tilde{\y}^{(n)},
	\end{equation}
	where $\circ$ denotes the entrywise product. By following the same reasoning that leads to \eqref{E:classical_equality_final} from \eqref{E:classical_equality}, we obtain that $\tilde{\Y}\tilde{\h}=\mathbf{0}$, where $\tilde{\h}:=[\tilde{\h}_1^{\top},\cdots,\tilde{\h}_M^{\top}]^{\top}$. Since $\tilde{\h}^{(m)} = \mathbf{\Psi}^{(m)}\h^{(m)}$ [cf. (\ref{E:output_f})], we have that $\tilde{\h}=\mathbf{\Psi}\h$ and the result holds.
\end{myproof}

It follows from Proposition \ref{Prop:multi} that, in the noise-free case we can estimate the filter coefficients $\h$ by solving \eqref{E:multi_basic}.
Notice that the outputs from multiple filters are correlated since they are excited by the same input, and the basic idea here is to take advantage of this correlation [cf. \eqref{E:multi_pair}]. 
Proposition \ref{Prop:multi} also reveals that the solution to \eqref{E:multi_basic} is identifiable if and only if $\rank(\tilde{\Y}\mathbf{\Psi})=\sum_{m=1}^M L_m-1$.

Next, we will give an interpretation of \eqref{E:multi_basic} in polynomial form which can help get further insights on the identifiability of the solution.
To this end, let us define the polynomial
\begin{equation}\label{E:polinomials_per_channel}
p^{(m)}(z) := \sum_{l=0}^{L_m-1}h_l^{(m)}z^l
\end{equation}
associated with $\h^{(m)}$ for $m=1,...,M$.
Moreover, we define the index set $\Omega_1\subseteq \{1,\cdots,N\}$ as the largest possible set such that for all $i\in\Omega_1$, $\tilde{x}_i\neq 0$ and every $\lambda_i$ is distinct.
Note that $|\Omega_1|$ equals the number of (distinct) graph frequencies contained in the input and thus reflects its spectral richness, which depends on both the input and the GSO.
The following relation can be shown between the polynomials $\hat{p}^{(m)}(z)$ associated with a generic solution $\hat{\h}$ to \eqref{E:multi_basic} and the polynomials $p^{(m)}(z)$ associated with the true filter coefficients ${\h}$.

\begin{myproposition}\label{Prop:multi_poly}
	The polynomials $\{\hat{p}^{(m)}(z)\}_{m=1}^M$ defined as in \eqref{E:polinomials_per_channel} associated with the solution $\hat{\h}$ to \eqref{E:multi_basic} satisfy
\begin{equation}\label{E:multi_poly}
p^{(m)}(\lambda_i){\hat{p}}^{(n)}(\lambda_i) = p^{(n)}(\lambda_i){\hat{p}}^{(m)}(\lambda_i)
\end{equation}
for all $1\leq m<n\leq M$ and $i\in\Omega_1$.
\end{myproposition}
\begin{myproof}
	First notice that the $i$th entry of $\tilde{\h}^{(m)}$ equals $p^{(m)}(\lambda_i)$. Thus, from (\ref{E:multi_pair}) we have
	\begin{equation}\label{E:multi_known_suffi_proof_010}
		p^{(m)}(\lambda_i){\hat{p}}^{(n)}(\lambda_i)\tilde{x}_i = p^{(n)}(\lambda_i){\hat{p}}^{(m)}(\lambda_i)\tilde{x}_i
	\end{equation}
for all $1\leq m<n\leq M$ and $i=1,\cdots,N$. According to the definition of $\Omega_1$, the result follows.
\end{myproof}

Proposition~\ref{Prop:multi_poly} relates the solution $\hat\h$ to \eqref{E:multi_basic} with polynomials $\{\hat{p}^{(m)}(z)\}_{m=1}^M$ that satisfy \eqref{E:multi_poly}.
Based on this, we obtain the sufficient conditions and necessary conditions for identifiability of the solution to \eqref{E:multi_basic} and state them in Theorems \ref{Th:multi_sufficient} and \ref{Th:multi_necessary}, respectively.

\begin{mytheorem}\label{Th:multi_sufficient}
The solution $\hat{\h}$ to (\ref{E:multi_basic}) can be identified uniquely (up to a scalar multiple) if:\\
i) $|\Omega_1|\geq L_{\max}+L_{\min}-1$, where $L_{\max}$ and $L_{\min}$ are the maximum and minimum values in $\{L_m\}_{m=1}^M$, and\\
ii) There does not exist a root shared by all the polynomials $\{p^{(m)}(z)\}_{m=1}^M$.
\end{mytheorem}

\begin{myproof}
We will show that under conditions \emph{i)} and \emph{ii)} it must be that $\hat{\h}=\alpha\h$.
Choose the index $m$ as one satisfying $L_m=L_{\min}$ and fix an arbitrary $n\neq m$.
It follows from Proposition~\ref{Prop:multi_poly} that the polynomial $p_{mn}(z) = p^{(m)}(z){\hat{p}}^{(n)}(z) - p^{(n)}(z){\hat{p}}^{(m)}(z)$ has at least $|\Omega_1|$ roots given by $\{\lambda_i\}_{i\in\Omega_1}$.
However, the degree of $p_{mn}(z)$ is at most $L_{\max}+L_{\min}-2$, thus, from condition \emph{i)} we have that $p_{mn}(z)=0$ for all $z\in\mathbb{C}$. 
Equivalently, we have that 
	\begin{equation}\label{E:multi_known_suffi_proof_020}
		p^{(m)}(z)\hat{p}^{(n)}(z) = p^{(n)}(z)\hat{p}^{(m)}(z)
	\end{equation}
for all $z\in\mathbb{C}$. It follows that each root $z_0$ of $p^{(m)}(z)$ -- i.e., a zero in the left-hand side of (\ref{E:multi_known_suffi_proof_020}) -- must be a root of $\hat{p}^{(m)}(z)$, since by condition \emph{ii)} there must exist an $n^{\ast}$ such that $p^{(n^{\ast})}(z_0)\neq 0$, and we know that $p^{(n^{\ast})}(z_0)\hat{p}^{(m)}(z_0)=0$. 
If $p^{(m)}(z)$ has distinct roots, we obtain that $\hat{p}^{(m)}(z)=\alpha p^{(m)}(z)$ for some scalar $\alpha$ since $\hat{p}^{(m)}(z)$ and $p^{(m)}(z)$ have the same degree.\footnote{The proof for the case of repeated roots is given in Appendix \ref{A:repeated_roots}.}
Finally, replacing this equality in (\ref{E:multi_known_suffi_proof_020}) we obtain that $\hat{p}^{(n)}(z)=\alpha p^{(n)}(z)$ for all arbitrary $n$, and thus $\hat\h=\alpha\h$.
\end{myproof}

\begin{mytheorem}\label{Th:multi_necessary}
The solution $\hat{\h}$ to (\ref{E:multi_basic}) is unidentifiable if:\\
i) $|\Omega_1|< L_{\max}$, or\\
ii) There exists a root shared by all the polynomials $\{p^{(m)}(z)\}_{m=1}^M$.
\end{mytheorem}

\begin{myproof}
We will first show that, under condition \emph{i)}, the filters are not identifiable even if the input is known.
The necessary condition of this problem should also be that of the blind identification problem since the latter one has no knowledge of the input.
If $\x$ is known, the $m$th filter can be found by solving $\diag(\tilde{\x})\mathbf{\Psi}^{(m)}\hat{\h}^{(m)}=\tilde{\y}^{(m)}$ [cf. (\ref{E:output_f})]. 
If $\tilde{x}_i=0$, the entries in the $i$th row of $\diag(\tilde{\x})\mathbf{\Psi}^{(m)}$ are all zero. 
If $\lambda_i=\lambda_j$, the $i$th row and the $j$th row of $\diag(\tilde{\x})\mathbf{\Psi}^{(m)}$ are linearly dependent. 
Hence, under condition \emph{i)}, the row rank of $\diag(\tilde{\x})\mathbf{\Psi}^{(m)}$ is less than $L_{\max}$, and thus the filter with the largest order cannot be uniquely identified.
For condition \emph{ii)}, if there is a common root $z_0$ shared by all the polynomials $\{p^{(m)}(z)\}_{m=1}^M$, we can write $p^{(m)}(z)=q^{(m)}(z)(z-z_0)$ for all $m$. In this case, the filter coefficients associated with the polynomials $q^{(m)}(z)(z-z'_0)$ for any choice of $z'_0$ also solve (\ref{E:multi_basic}).
\end{myproof}

Theorems \ref{Th:multi_sufficient} and \ref{Th:multi_necessary} reveal how the identifiability of the solution to \eqref{E:multi_basic} depends on $|\Omega_1|$ and $\{p^{(m)}(z)\}_{m=1}^M$, which in fact characterize the spectral richness of the input and the filter coefficients, respectively. These theorems successfully parallel the classical results in \cite[Thm. 1 and 2]{xu_1995_least}.
Notice that in \cite{xu_1995_least} multiple channels are assumed to have the same order, while here we do not make such assumption and consider the more general case.
In addition, $|\Omega_1|$ depends on both the input and the underlying graph, while the richness of the input in \cite{xu_1995_least} is 
defined as the linear complexity of the input sequence and exclusively determined by the input since there is no underlying graph in the classical setting. 
By Theorems \ref{Th:multi_sufficient} and \ref{Th:multi_necessary}, if there does not exist a root shared by all the polynomials $\{p^{(m)}(z)\}_{m=1}^M$, 
the filters are identifiable if $|\Omega_1|\geq L_{\max}+L_{\min}-1$, and they cannot be uniquely identified if $|\Omega_1|<L_{\max}$.
To tackle the case where $L_{\max}\leq |\Omega_1|<L_{\max}+L_{\min}-1$, the sufficient and necessary identifiability condition is given in Theorem~\ref{Th:multi_sufficient_necessary} in which the spectral richness of the input and the polynomials associated with the filter coefficients cannot be decoupled.

\begin{mytheorem}\label{Th:multi_sufficient_necessary}
Let $z_1,z_2,\cdots,z_{|\Omega_1|}$ denote the entries in $\pmb{\lambda}_{\Omega_1}$.
Define $\p(z_i):=[p^{(1)}(z_i),p^{(2)}(z_i),\cdots,p^{(M)}(z_i)]^{\top}$.
Let $\Z(z_i)$ denote a block diagonal matrix which contains $M$ main-diagonal blocks with the $m$th one being $[1,z_i,\cdots,z_i^{L_m-1}]$.
Then, the solution $\hat{\h}$ to (\ref{E:multi_basic}) can be identified uniquely (up to a scalar multiple) if and only if the matrix
	\begin{equation}\label{E:multi_matrix}
		\left[\begin{matrix}
		\p(z_1) & & \mathbf{0} & \Z(z_1) \\
            	&\ddots& & \vdots\\
		\mathbf{0} & & \p(z_{|\Omega_1|}) & \Z(z_{|\Omega_1|})
		\end{matrix}\right]
	\end{equation}
has a one-dimensional null space.
\end{mytheorem}

\begin{myproof}
This proof is given in Appendix~\ref{A:proof_multi_known}.
\end{myproof}

Notice that the matrix in (\ref{E:multi_matrix}) has $|\Omega_1|+\sum_{m=1}^M L_m$ columns and $M|\Omega_1|$ rows. 
If it has a one-dimensional null space, we should have $M|\Omega_1|\geq |\Omega_1|+\sum_{m=1}^M L_m-1$.
Hence, Theorem \ref{Th:multi_sufficient_necessary} implies that the solution is not identifiable if $(M-1)|\Omega_1|<\sum_{m=1}^M L_m-1$.
Moreover, as shown in Appendix \ref{A:proof_multi_known}, the vector 
\begin{equation}\label{E:solution}
[ \underbrace{1,1,\cdots,1}_{|\Omega_1|},-\h^{\top}]^{\top}
\end{equation}
is always in the null space of the matrix in \eqref{E:multi_matrix} no matter what the values of $z_1,\cdots,z_{|\Omega_1|}$ are.
Thus, Theorem~\ref{Th:multi_sufficient_necessary} can be interpreted as stating that the frequencies contained in the input (i.e., $\pmb{\lambda}_{\Omega_1}$) cannot allow other vectors -- independent of that in \eqref{E:solution} -- to be in the null space of the matrix in \eqref{E:multi_matrix}.
The frequencies contained in the input are the counterpart to the input modes in the classical setting \cite{xu_1995_least}.

Thus far, we have established identifiability conditions for recovery in the absence of noise. 
However, in the practical case where there exists noise in the observations, one can estimate $\h$ by solving the following simple least-squares problem
\begin{equation}
\hat{\h} = \min_{\|\h\|_2=1}  \|\tilde{\Y}\mathbf{\Psi}\h\|_2^2,
\label{eq:multi_known_noise}
\end{equation}
where the constraint is added in order to eliminate the scalar ambiguity. 

\subsection{Unknown Filter Orders}\label{SS:multi_unknown}

When the filter orders are unknown, the approach proposed in Section \ref{SS:multi_known} is no longer feasible since it requires knowledge of $\{L_m\}_{m=1}^M$ to build matrix $\mathbf{\Psi}$.
A workaround for this issue is to overshoot the filter orders, and then incorporate a sparsity based regularizer that promotes shorter filters whenever they can appropriately explain the observed outputs.
To be precise, we assume that the order of the $m$th filter is $Q_m$, where we ensure that $Q_m\geq L_m$ for all $1\leq m\leq M$ by selecting large enough $Q_m$ as guided by available domain knowledge. 
We then define a series of Vandermonde matrices $\{\mathbf{\Theta}^{(m)}\}_{m=1}^M$, where $\mathbf{\Theta}^{(m)}$ is of dimension $N\times Q_m$ and $[\mathbf{\Theta}^{(m)}]_{ij} := \lambda_i^{j-1}$.
We further define a block diagonal matrix $\mathbf{\Theta}$ whose $M$ main-diagonal blocks are respectively given by $\{\mathbf{\Theta}^{(m)}\}_{m=1}^M$.
Notice that $\mathbf{\Theta}=\mathbf{\Psi}$ if $Q_m=L_m$ for all $m$.
Based on the true filter coefficients $\h^{(m)}$, define the zero-padded vectors $\bar{\h}^{(m)}:=[\h^{(m)\top},\mathbf{0}^{\top}_{(Q_m-L_m)\times 1}]^{\top}$ and $\bar{\h}:=[\bar{\h}^{(1)\top},\cdots,\bar{\h}^{(M)\top}]^{\top}$.
Clearly, $\mathbf{\Theta}^{(m)}\bar{\h}^{(m)} = \mathbf{\Psi}^{(m)}\h^{(m)}$ and thus $\mathbf{\Theta}\bar{\h} = \mathbf{\Psi}\h$. 
Hence, it follows from Proposition \ref{Prop:multi} that
	\begin{equation}\label{E:multi_basic_unknown}
		\tilde{\mathbf{Y}}\mathbf{\Theta}\bar\h=\mathbf{0}.
	\end{equation}
We can estimate $\bar{\h}$ by solving \eqref{E:multi_basic_unknown} as the approach proposed in Section \ref{SS:multi_known}.	
But it becomes harder to guarantee the identifiability of the solution since the number of unknown parameters to be determined has been increased due to overshooting the filter orders.
Hence, we propose to estimate $\bar\h$ by solving the following convex optimization problem in which we leverage a priori information, i.e., that the solution is sparse, 
\begin{equation}\label{E:l1_analysis_problem}
\hat{\bar{\bbh}} = \argmin_{\bar{\bbh}} \| \bbDelta \bar{\bbh} \|_1 \quad \text{s.t. } \tilde{\mathbf{Y}} \bbTheta \bar{\bbh} = \mathbf{0}, \,\, \bar{h}_1 = 1,
\end{equation}
where $\mathbf{\Delta}$ is a diagonal matrix that contains positive predefined weights.
When we choose $\mathbf{\Delta}=\I$, (\ref{E:l1_analysis_problem}) seeks among all non-trivial solutions to $\tilde{\Y}\mathbf{\Theta}\bar{\h}=\mathbf{0}$ for the sparsest one where the $\ell_1$ norm has been used as a convex surrogate of the non-convex $\ell_0$ pseudo-norm. 
Moreover, given that the true $\bar\h$ is formed by concatenating $M$ zero-padded vectors $\{\bar{\h}^{(m)}\}_{m=1}^M$, when solving (\ref{E:l1_analysis_problem}) we seek to strongly promote zeros for the filters coefficients associated with larger degrees in every $\bar{\h}^{(m)}$.
This can be achieved by setting increasing weights in each $\mathbf{\Delta}^{(m)}$, where $\mathbf{\Delta}^{(m)}\in\mathbb{R}^{Q_m\times Q_m}$ is the $m$th main-diagonal block matrix in $\mathbf{\Delta}$.
Indeed, the experiment results in Section \ref{S:num_exp} will show that there is a clear advantage associated with the consideration of weight matrices $\bbDelta$ different from the identity matrix. 
Lastly, the second constraint in (\ref{E:l1_analysis_problem}) is simply enforced to avoid the trivial solution $\hat{\bar{\h}}=\mathbf{0}$. 
Given that the recovery is always considered up to a scalar multiple, this constraint is only making the mild assumption that $h^{(1)}_1\neq 0$ for the true filters.

In order to further simplify (\ref{E:l1_analysis_problem}),
let us write $\tilde{\Y}\mathbf{\Theta}=[\b, \mathbf{\Phi}]$ where $\b$ is the first column of $\tilde{\Y}\mathbf{\Theta}$ and $\mathbf{\Phi}$ consists of the remaining columns.
We denote by $\bar{\h}_{(-1)}$ the vector obtained by dropping the first entry of $\bar{\h}$ and by $\mathbf{\Delta}_{(-1)}$ the matrix obtained by dropping the first column and the first row of $\mathbf{\Delta}$.
Then, (\ref{E:l1_analysis_problem}) can be rewritten as
\begin{equation}\label{E:l1_analysis_problem2}
\hat{\bar{\bbh}}_{(-1)} = \argmin_{\bar{\bbh}_{(-1)}} \| \bbDelta_{(-1)} \bar{\bbh}_{(-1)} \|_1 \,\, \text{ s.t. } \bbPhi \bar{\bbh}_{(-1)} = -\bbb.
\end{equation}
Problem~\eqref{E:l1_analysis_problem2} is in the form of an $\ell_1$-analysis model \cite{zhang_2016_one}.
Next, in Theorem \ref{Th:multi_unknown_sufficient} we show the sufficient conditions under which the solution to \eqref{E:l1_analysis_problem2} -- or, equivalently, to \eqref{E:l1_analysis_problem} -- is unique and coincides with the true filter coefficients $\bar{\bbh}$.
We denote by $\mathcal{I}=\mathrm{supp}({\bar\h}_{(-1)})\subset\{1,\cdots,\sum_{m=1}^{M}Q_m -1\}$ the set of indices of the non-zero true coefficients (after dropping $\bar{h}_1$), and by $\mathcal{I}^c$ its complement. 
With this notation in place, the following result can be shown.

\begin{mytheorem}\label{Th:multi_unknown_sufficient}
	The solution $\hat{\bar\h}$ to \eqref{E:l1_analysis_problem} coincides with the true $\bar{\bbh}$ if the following two conditions hold:\\
	i) $\rank(\mathbf{\Phi}_{\mathcal{I}} )= |\mathcal{I}|$, and\\
	ii) There exists a constant $\delta>0$ such that 
		$$\xi := \|\mathbf{I}_{\ccalI^c}^\top(\delta^{-2}\bbDelta_{(-1)}^{-1}\mathbf{\Phi}^\top\mathbf{\Phi}\bbDelta_{(-1)}^{-1}+\mathbf{I}_{\ccalI^c}\mathbf{I}_{\ccalI^c}^\top)^{-1}\mathbf{I}_{\mathcal{I}}\|_{\infty} <1.$$
\end{mytheorem}

\begin{myproof}
See Appendix \ref{A:proof_multi_unknown}. 
\end{myproof}

In Theorem~\ref{Th:multi_unknown_sufficient}, condition \emph{i)} requires the null space of matrix $\mathbf{\Phi}_{\mathcal{I}}$ to have dimension zero. Otherwise, there would exist a non-zero vector $\r$ satisfying $\mathbf{\Phi}_{\mathcal{I}}\r_{\mathcal{I}}=\mathbf{0}$ and $\r_{\mathcal{I}^c}=\mathbf{0}$ such that $\mathbf{\Phi}(\bar{\h}_{(-1)}+\r)=\mathbf{\Phi}\bar{\h}_{(-1)}$. This would imply the existence of a nonempty interval $\mathcal{R} = [\bar{\h}_{(-1)}-\alpha\r,\bar{\h}_{(-1)}+\alpha\r]$ where $\alpha>0$ is sufficiently small so that $\mathbf{\Phi}\bar\r=\mathbf{\Phi}\bar{\h}_{(-1)}$ and $\|\mathbf{\Delta}_{(-1)}\bar\r\|_1$ is linear for $\bar{\r}\in\mathcal{R}$.
Hence, $\bar{\h}_{(-1)}$ could not be the unique solution of~\eqref{E:l1_analysis_problem2}.
Condition \emph{ii)} is derived from the construction of a strictly-complementary dual certificate which is able to certify the optimality of $\bar{\h}_{(-1)}$.

When there exists noise in the observed outputs, (\ref{E:l1_analysis_problem2}) can be adapted to 
\begin{equation}\label{E:multi_unknown_noise}
\hat{\bar{\bbh}}'_{(-1)} \!=\! \argmin_{\bar{\bbh}_{(-1)}} \| \bbDelta_{(-1)} \bar{\bbh}_{(-1)} \|_1  \text{ s.t.}
\| \bbPhi \bar{\bbh}_{(-1)} +\bbb\|_2 \!\leq\!\epsilon,
\end{equation}
where $\epsilon$ depends on the noise strength. 
We denote the maximum and minimum diagonal entries of $\mathbf{\Delta}_{(-1)}$ as $\Delta_{\max}$ and $\Delta_{\min}$ respectively, which also correspond to the maximum and minimum singular values of $\mathbf{\Delta}_{(-1)}$.
And we assume that $\mathbf{\Delta}_{(-1)}$ is normalized to satisfy $\Delta_{\max}=1$.
With this notation and assumption, the following theorem shows the result on robust recovery of multiple filters.

\begin{mytheorem}\label{Th:multi_unknown_noise}
For arbitrary noise $\w$, let $\mathbf{\Phi}\bar{\h}_{(-1)}+\b=\w$ and $\epsilon=\|\w\|_2$. 
If the conditions stated in Theorem \ref{Th:multi_unknown_sufficient} are met, then every minimizer $\hat{\bar\h}'_{(-1)}$ of problem (\ref{E:multi_unknown_noise}) satisfies
\begin{equation}
	\left\|\pmb{\Delta}_{(-1)}\left(\hat{\bar\h}'_{(-1)}-\bar\h_{(-1)}\right)\right\|_1\leq C\epsilon, \quad C=2C_1+2C_2\sqrt{C_3},\nonumber
\end{equation}
where
\begin{small}
\begin{equation}
C_1 \!=\! \frac{\sqrt{|\mathcal{I}|}}{\sigma_{\min}(\mathbf{\Phi}_{\mathcal{I}})}, C_2\!=\!\frac{1\!+\!\frac{\Delta_{\max}}{\Delta_{\min}}C_1\|\mathbf{\Phi}\|_2}{1-\psi}, C_3 \!=\! \|\mathbf{\Phi}^{\dag}\mathbf{\Delta}\|_2^2 \cdot \sum_{m=1}^M Q_m, \nonumber
\end{equation}
\end{small}
and $\sigma_{\min}(\cdot)$ denotes the minimum singular value of the argument matrix.
\end{mytheorem}

\begin{myproof}
See Appendix \ref{A:proof_multi_noise}.
\end{myproof}

Theorem~\ref{Th:multi_unknown_noise} quantifies the effect of the noise in recovering the filter coefficients. More precisely, the distance between the recovered vector of filter coefficients and the true one is upper bounded by the noise strength $\epsilon$ times a constant, which depends on $\mathbf{\Phi}$, $\mathbf{\Delta}$, $\mathcal{I}$, and the assumed filter orders $\{Q_m\}_{m=1}^M$.

\section{Estimation of a Single Network Process}\label{S:single_process}

The assumption that multiple network processes are driven by a common input might not hold in some practical scenarios.  
This section considers a more pragmatic setting where the outputs $\{\y^{(m)}\}_{m=1}^M$ are sequentially sampled from a single network process at different points in time and, thus, the same input assumption is naturally satisfied.
For this case, the graph filters defined in Section \ref{SS:ps} are dependent and partially share common coefficients. 
More precisely, assuming that $L_1<\cdots<L_M$, the filters can be rewritten as $\mathbf{H}^{(m)}=\sum_{l=0}^{L_m-1}h^{(m)}_l\S^l$.
To eliminate redundancy in $\{\h^{(m)}\}_{m=1}^M$ and concentrate on the newly added coefficients in each filter, we define $\d^{(1)}=\h^{(1)}$ and $\d^{(m)}$ as the vector collecting the last $L_m-L_{m-1}$ elements of $\h^{(m)}$ for $2\leq m\leq M$, then we can write $\h^{(m)}=[\h^{(m-1)\top},\d^{(m)\top}]^{\top}$ and $\h^{(m)}=[\d^{(1)\top},\cdots,\d^{(m)\top}]^{\top}$.
The goal is to recover $\{\d^{(m)}\}_{m=1}^M$.

\subsection{Known Filter Orders}\label{SS:single_known}
With known filter orders, we can define the following matrix
\begin{equation}
\bar{\mathbf{\Psi}} := \left[\begin{matrix}
		\mathbf{\Psi}^{(1)} \quad \mathbf{0}_{N\times (L_M-L_1)} \\
		\mathbf{\Psi}^{(2)} \quad \mathbf{0}_{N\times (L_M-L_2)} \\
\vdots\\
\mathbf{\Psi}^{(M)}
\end{matrix}\right],
\end{equation}
where the definitions of $\{\mathbf{\Psi}^{(m)}\}_{m=1}^M$ are given in Section \ref{SS:multi_known}. From (\ref{E:multi_basic}), it follows that
\begin{equation}\label{E:single_basic}
\tilde{\Y}\bar{\mathbf{\Psi}}\d = \mathbf{0},
\end{equation}
where $\d:=[\d^{(1)\top},\cdots,\d^{(M)\top}]^{\top}$ and thus $\d=\h^{(M)}$.
It can be observed from (\ref{E:single_basic}) that we can recover the true $\d$ if and only if $\rank(\tilde{\Y}\bar{\mathbf{\Psi}})=L_M-1$.
In order to express (\ref{E:single_basic}) in polynomial form so that we can have a better understanding of the relations between this problem and the one in Section \ref{S:multi_processes}, let us define the polynomials 
\begin{subequations}\label{E:polyd}
\begin{alignat}{2}
d^{(1)}(z)& := \sum\nolimits_{l=0}^{L_1-1}d^{(1)}_l z^l, \\
d^{(m)}(z)& := \sum\nolimits_{l=0}^{L_m-L_{m-1}-1} d^{(m)}_l z^l, \quad 2\leq m\leq M,
\end{alignat}
\end{subequations}
whose coefficients are respectively given by $\{\d^{(m)}\}_{m=1}^M$.
Notice that, $d^{(1)}(z)=p^{(1)}(z)$ and $d^{(m)}(z)=z^{-L_{m-1}}(p^{(m)}(z)-p^{(m-1)}(z))$ for $2\leq m\leq M$. 
Moreover, define the index set $\Omega_2\in\{1,\cdots,N\}$ as the largest possible set satisfying that $\tilde{x}_i\neq 0$ for all $i\in\Omega_2$ and the eigenvalues $\lambda_i$ indexed by $i\in\Omega_2$ are non-zero and non-repeated. 
Note that $\Omega_2\subseteq\Omega_1$ (cf. the definition of $\Omega_1$ in Section~\ref{SS:multi_known}), due to the added requirement of non-zero eigenvalues.

\begin{myproposition}\label{Prop:single_poly}
The polynomials $\{\hat{d}^{(m)}(z)\}_{m=1}^M$ defined as in \eqref{E:polyd} associated with the solution $\hat\d$ to \eqref{E:single_basic} satisfy
\begin{equation}\label{E:single_poly}
d^{(m)}(\lambda_i){\hat{d}}^{(n)}(\lambda_i) = d^{(n)}(\lambda_i){\hat{d}}^{(m)}(\lambda_i)
\end{equation}
for all $1\leq m<n\leq M$ and $i\in\Omega_2$.
\end{myproposition}

\begin{myproof}
Define $d_{mn}(z) := z^{-L_m}(p^{(n)}(z)-p^{(m)}(z))$.
Notice that $d_{mn}(z)=d^{(n)}(z)$ when $n=m+1$.
From Proposition~\ref{Prop:multi_poly}, we have that
\begin{align}\label{single_poly_proof_2}
p^{(m)}(\lambda_i)&\left( {\hat{p}}^{(m)}(\lambda_i) + \lambda_i^{L_m}\hat{d}_{mn}(\lambda_i) \right) \nonumber\\
&=\left( {{p}}^{(m)}(\lambda_i) + \lambda_i^{L_m}{d}_{mn}(\lambda_i) \right){\hat{p}}^{(m)}(\lambda_i),
\end{align}
for all $1\leq m< n\leq M$ and $i\in\Omega_2$, and it follows that
\begin{equation}\label{single_poly_proof_3}
p^{(m)}(\lambda_i)\hat{d}_{mn}(\lambda_i)  = {d}_{mn}(\lambda_i){\hat{p}}^{(m)}(\lambda_i).
\end{equation}
When $n>m+1$, replace $n$ in (\ref{single_poly_proof_3}) with $n-1$ and we get 
\begin{equation}\label{single_poly_proof_4}
p^{(m)}(\lambda_i)\hat{d}_{m(n-1)}(\lambda_i)  = {d}_{m(n-1)}(\lambda_i){\hat{p}}^{(m)}(\lambda_i).
\end{equation} 
Subtracting \eqref{single_poly_proof_4} from \eqref{single_poly_proof_3}, we obtain that
\begin{equation}\label{single_poly_proof_5}
p^{(m)}(\lambda_i)\hat{d}^{(n)}(\lambda_i)  = {d}^{(n)}(\lambda_i){\hat{p}}^{(m)}(\lambda_i),
\end{equation}
where we leveraged the fact that $d_{mn}(z)-d_{m(n-1)}(z) = d^{(n)}(z)z^{L_{n-1}-L_m}$. 
For $n=m+1$, we can directly obtain \eqref{single_poly_proof_5} from \eqref{single_poly_proof_3}.
Hence, \eqref{single_poly_proof_5} holds for all $n>m$.
For $m>1$, replace $m$ in (\ref{single_poly_proof_5}) with $m-1$ and we have that
\begin{equation}\label{single_poly_proof_6}
p^{(m-1)}(\lambda_i)\hat{d}^{(n)}(\lambda_i)  = {d}^{(n)}(\lambda_i){\hat{p}}^{(m-1)}(\lambda_i).
\end{equation}
Subtract \eqref{single_poly_proof_6} from \eqref{single_poly_proof_5} and we obtain \eqref{E:single_poly}.
When $m=1$, we can directly obtain \eqref{E:single_poly} from \eqref{single_poly_proof_5}. 
Hence, \eqref{E:single_poly} holds for all $1\leq m<n\leq M$ and $i\in\Omega_2$.
\end{myproof}

First notice that the results in Section \ref{S:multi_processes} including Proposition \ref{Prop:multi_poly} are still valid for the problem discussed in this section. 
However, the fact that the filters share part of the coefficients implies that the results in Section~\ref{S:multi_processes} are not tight enough for this case.
Proposition~\ref{Prop:single_poly} leverages the additional structure presented here, tailoring Proposition~\ref{Prop:multi_poly} to the estimation problem of a single network process.
Moreover, here we consider $\Omega_2$ instead of $\Omega_1$ since \eqref{E:multi_poly} always holds when $\lambda_i=0$ for the single network process scenario where $p^{(m)}(0)=h^{(M)}_1$ for all $1\leq m\leq M$.
Since Proposition \ref{Prop:single_poly} has the same form as Proposition \ref{Prop:multi_poly} from where Theorems \ref{Th:multi_sufficient}-\ref{Th:multi_sufficient_necessary} are developed, we can use similar methods to obtain identification conditions for problem \eqref{E:single_basic} based on Proposition \ref{Prop:single_poly}.
Theorems \ref{Th:single_sufficient} and \ref{Th:single_sufficient_necessary} can be proved in the same way as shown in the proofs for Theorems \ref{Th:multi_sufficient} and \ref{Th:multi_sufficient_necessary} respectively, and thus their proofs are omitted below.
By contrast, we include the proof of the first condition in Theorem~\ref{Th:single_necessary} since it does not directly follow from the proof of Theorem~\ref{Th:multi_necessary}.

\begin{mytheorem}\label{Th:single_sufficient}
The solution $\hat{\d}$ to (\ref{E:single_basic}) can be identified uniquely (up to a scalar multiple) if:\\
i) $|\Omega_2|\geq \bar{L}_{\max}+\bar{L}_{\min}-1$, where $\bar{L}_{\max}$ and $\bar{L}_{\min}$ are the maximum and minimum values in the set $\{L_1,L_2-L_1,\cdots,L_M-L_{M-1}\}$, and\\
ii) There does not exist a root shared by all the polynomials $\{d^{(m)}(z)\}_{m=1}^M$.
\end{mytheorem}

\begin{mytheorem}\label{Th:single_necessary}
The solution $\hat{\d}$ to (\ref{E:single_basic}) is unidentifiable if:\\
i) $\min\{L_1,|\Omega_1|\}+\sum_{m=2}^M\min\{L_m-L_{m-1},|\Omega_2|\}<L_M$, or\\
ii) There exists a root shared by all the polynomials $\{d^{(m)}(z)\}_{m=1}^M$.
\end{mytheorem}

\begin{myproof}
	Define a block diagonal matrix $\tilde{\X}$ which has $M$ main-diagonal blocks $\diag(\tilde{\x})$ and set $\tilde{\y}:=[\tilde{\y}^{(1)\top},\cdots,\tilde{\y}^{(M)\top}]^{\top}$. Then, all the equations $\diag(\tilde{\x})\mathbf{\Psi}^{(m)}\hat{\h}^{(m)}=\tilde{\y}^{(m)}$ for $1\leq m\leq M$ can be condensed in
\begin{equation}\label{E:single_necessary_proof_1}
\tilde{\X}\bar{\mathbf{\Psi}}\hat{\d}=\tilde{\y}.
\end{equation}
If the input is known, the solution $\hat{\d}$ can be found by solving (\ref{E:single_necessary_proof_1}). 
If (\ref{E:single_necessary_proof_1}) does not have a unique solution, neither does the blind identification problem.
Rewrite $\mathbf{\Psi}^{(M)}=[\mathbf{\Psi}_1^{(M)},\cdots,\mathbf{\Psi}_M^{(M)}]$ where $\mathbf{\Psi}_1^{(M)}$ collects the first $L_1$ columns, $\mathbf{\Psi}_2^{(M)}$ collects the following $L_2-L_1$ columns and so forth.
The matrix $\tilde{\X}\bar{\mathbf{\Psi}}$ can be transformed to a block diagonal matrix whose $M$ main-diagonal blocks are respectively given by $\{\diag(\tilde{\x})\mathbf{\Psi}_m^{(M)}\}_{m=1}^M$ via elementary row operations. 
Hence, we have
\begin{equation}
\rank(\tilde{\X}\bar{\mathbf{\Psi}}) = \sum_{m=1}^M \rank(\diag(\tilde{\x})\mathbf{\Psi}_m^{(M)}),
\end{equation}
where 
\begin{align}
\rank(\diag(\tilde{\x})\mathbf{\Psi}_1^{(M)}) &= \min\{L_1,|\Omega_1|\}, \nonumber\\
\rank(\diag(\tilde{\x})\mathbf{\Psi}_m^{(M)}) &= \min\{L_m-L_{m-1},|\Omega_2|\}, 2\leq m\leq M. \nonumber
\end{align}
Therefore, under condition \emph{i)}, we get that $\rank(\tilde{\X}\bar{\mathbf{\Psi}})<L_M$ and the solution to (\ref{E:single_necessary_proof_1}) cannot be identified uniquely.
The proof of condition \emph{ii)} is similar to that of condition \emph{ii)} in Theorem \ref{Th:multi_necessary}, and thus is omitted. 
\end{myproof}

Condition \emph{i)} in Theorem \ref{Th:single_necessary} implies that the solution to (\ref{E:single_basic}) is not identifiable if $|\Omega_1|<L_1$ or $|\Omega_2|<\max\limits_{2\leq m\leq M}\{L_m-L_{m-1}\}$.
Theorems \ref{Th:single_sufficient} and \ref{Th:single_necessary} respectively state the sufficient conditions and necessary conditions for the identifiability of the solution to \eqref{E:single_basic}, and the sufficient and necessary condition is given in the following theorem.

\begin{mytheorem}\label{Th:single_sufficient_necessary}
Let $\bar{z}_1,\bar{z}_2,\cdots,\bar{z}_{|\Omega_2|}$ denote the entries in $\pmb{\lambda}_{\Omega_2}$.
Define $\d(\bar{z}_i):=[d^{(1)}(\bar{z}_i),d^{(2)}(\bar{z}_i),\cdots,d^{(M)}(\bar{z}_i)]^{\top}$.
Let $\bar\Z(\bar{z}_i)$ denote a block diagonal matrix whose $M$ main-diagonal blocks are given by $[1,\bar{z}_i,\cdots,\bar{z}_i^{L_1-1}]$ and $[1,\bar{z}_i,\cdots,\bar{z}_i^{L_m-L_{m-1}-1}]$ for $2\leq m\leq M$.
Then, the solution $\hat{\d}$ to (\ref{E:single_basic}) can be identified uniquely (up to a scalar multiple) if and only if the matrix
	\begin{equation}\label{E:single_matrix}
		\left[\begin{matrix}
		\d(\bar{z}_1) & & \mathbf{0} & \bar\Z(\bar{z}_1) \\
            	&\ddots& & \vdots\\
		\mathbf{0} & & \d(\bar{z}_{|\Omega_2|}) & \bar\Z(\bar{z}_{|\Omega_2|})
		\end{matrix}\right]
	\end{equation}
has a one-dimensional null space.
\end{mytheorem}

Mimicking the discussion after Theorem~\ref{Th:multi_sufficient_necessary}, notice that the matrix in \eqref{E:single_matrix} has $|\Omega_2|+L_M$ columns and $M|\Omega_2|$ rows. If it has a one-dimensional null space, we should have $M|\Omega_2|\geq |\Omega_2|+L_M-1$, which implies that we cannot identify the filters uniquely if $(M-1)|\Omega_2|<L_M-1$.

\subsection{Unknown Filter Orders}\label{SS:single_unknown}

\begin{figure*}
	\centering
	\subfigure[]{
		\centering
		\includegraphics[width=0.31\textwidth]{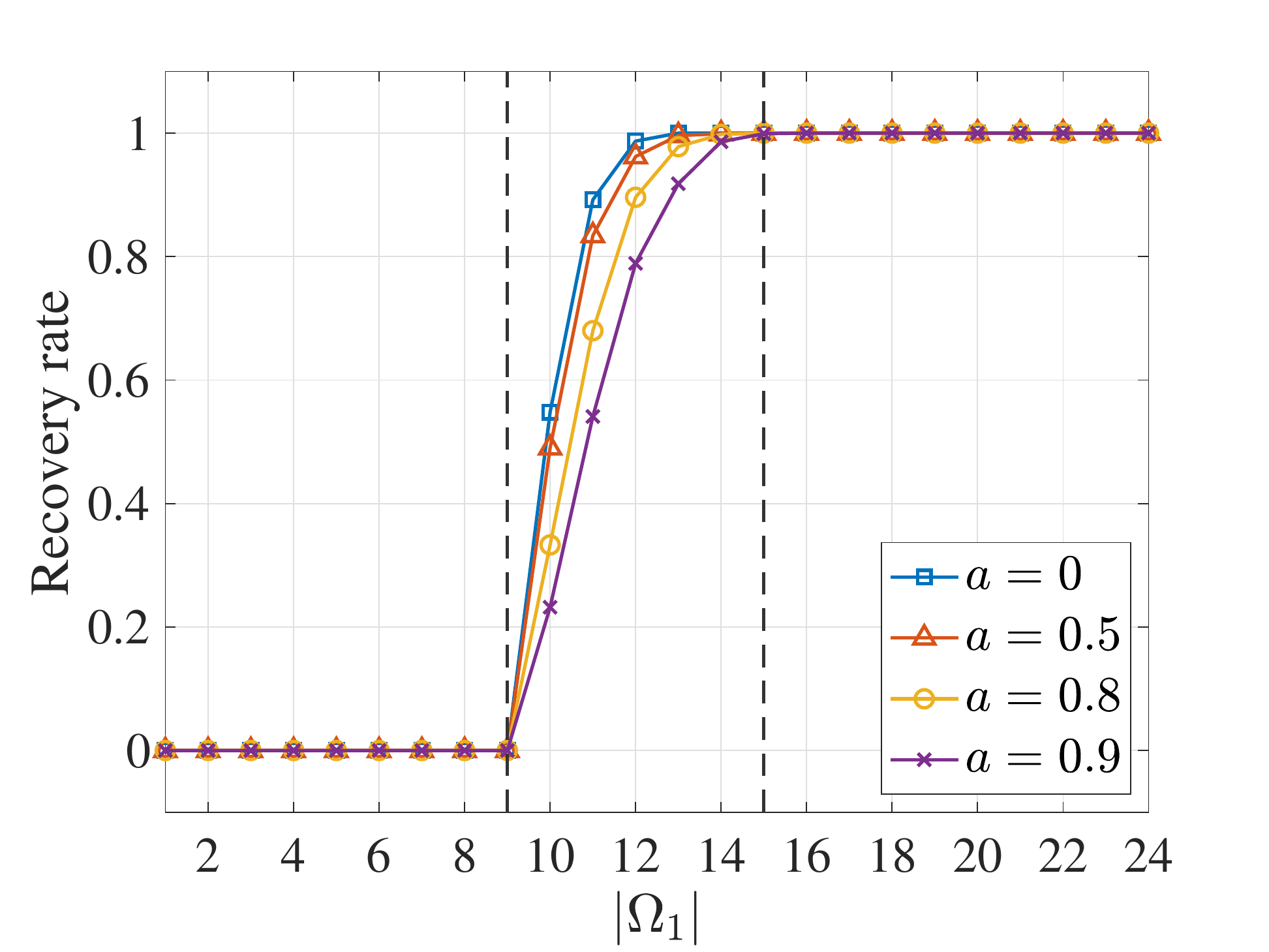}
		\label{Fig:verify_thm123}
	}	
	\subfigure[]{
		\centering
		\includegraphics[width=0.31\textwidth]{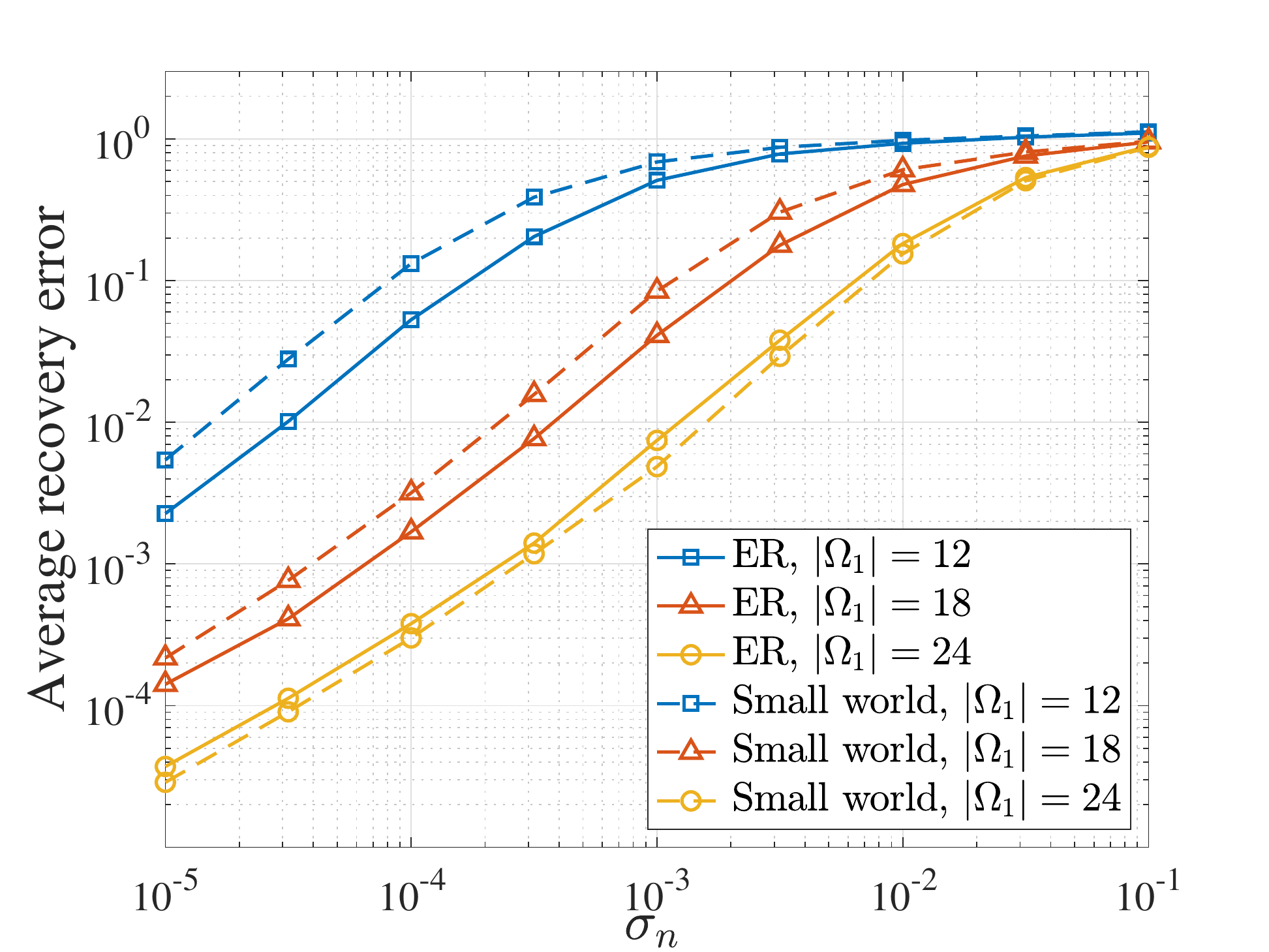}
		\label{Fig:influence_omega1}
	}
	\subfigure[]{
		\centering
		\includegraphics[width=0.31\textwidth]{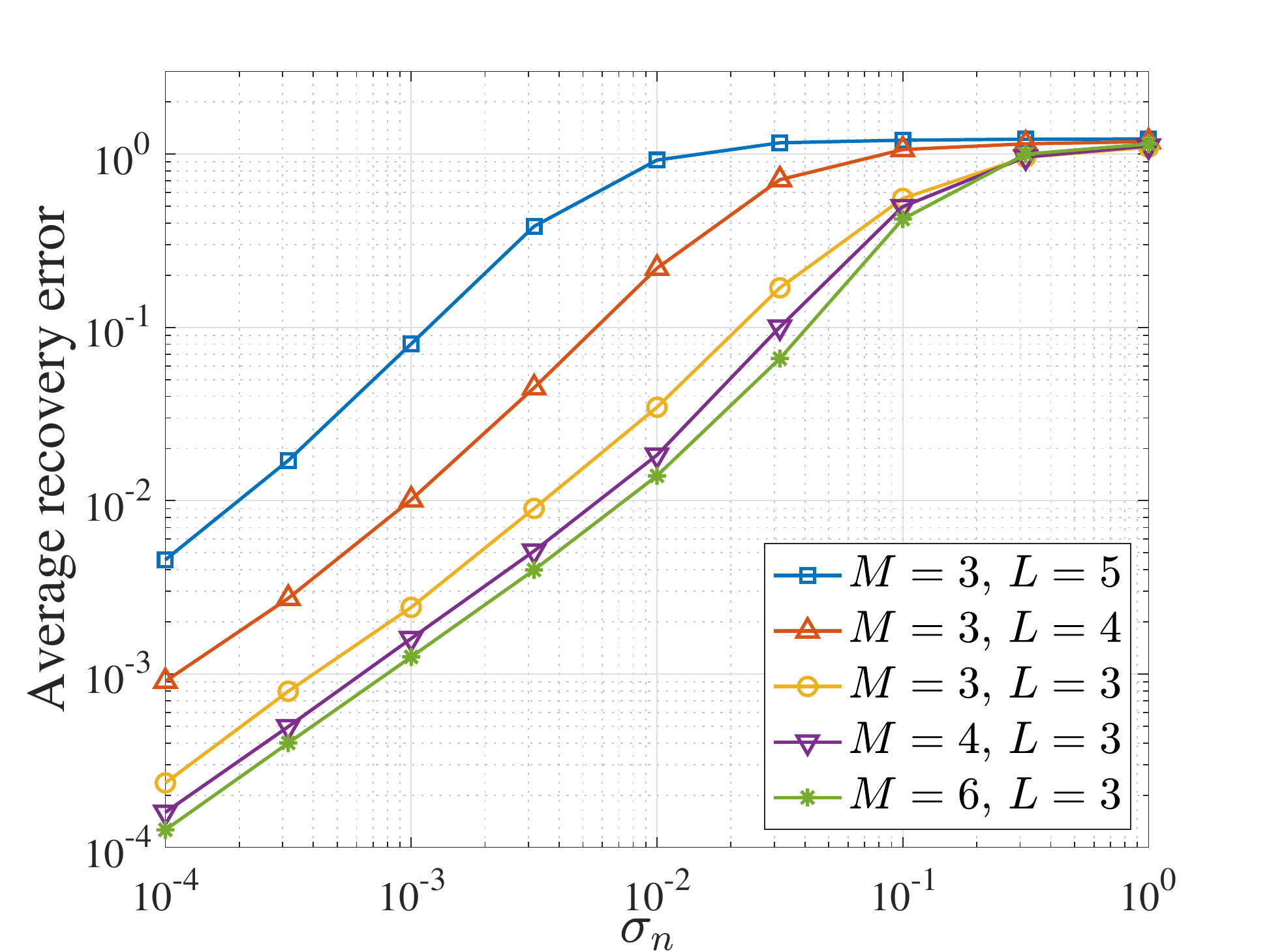}
		\label{Fig:multi_M_L}
	}	
	\vspace{-2mm}
	\caption{\small Estimation of multiple network processes when their filter orders are known. (a) Experimental validation of Theorems~\ref{Th:multi_sufficient}-\ref{Th:multi_sufficient_necessary}. For this setting ($M=5$ filters of order $L=8$), the filters are identifiable whenever $|\Omega_1|\geq 15$ and unidentifiable whenever $|\Omega_1|\leq 9$, as predicted by our theory. Higher correlations between the filters (larger values of $a$) make the recovery more challenging. (b) Average recovery error as a function of the noise level $\sigma$ for {different graph types and} different $|\Omega_1|$. Larger values of $|\Omega_1|$ facilitate recovery. (c) Average recovery error as a function of $\sigma$ for different numbers of filters $M$ and filter orders $L$. Larger $M$ and smaller $L$ result in better performance.}
	\vspace{-2mm}
\end{figure*}

In this section, we consider the case of unknown filter orders while we assume that outputs are sequentially sampled, i.e., the filter orders satisfy $L_1<\cdots<L_M$ although we do not know their exact values.
Similar to Section \ref{SS:multi_unknown}, we assume that the order of the $m$th filter is $Q_m$ and ensure that $Q_m\geq L_m$.
Define the following matrix
\begin{equation}
\bar{\mathbf{\Theta}}:=\left[\begin{matrix}
		\mathbf{\Theta}^{(1)} & \bb0 & \ldots & \bb0\\
		\mathbf{\Theta}^{(1)} & \mathbf{\Theta}^{(2)} & \ddots &\vdots\\
\vdots & & \ddots &\bb0\\
\mathbf{\Theta}^{(1)} & \mathbf{\Theta}^{(2)} & \cdots & \mathbf{\Theta}^{(M)}
\end{matrix}\right],
\end{equation}
where the definitions of $\{\mathbf{\Theta}^{(m)}\}_{m=1}^M$ are given in Section \ref{SS:multi_unknown}. 
Moreover, let us define $\bar{\d}^{(1)}:=[\d^{(1)\top},\mathbf{0}^{\top}_{(Q_1-L_1)\times 1}]^{\top}$ and $\bar{\d}^{(m)}:=[\mathbf{0}^{\top}_{L_{m-1}\times 1},\d^{(m)\top},\mathbf{0}^{\top}_{(Q_m-L_m)\times 1}]^{\top}$ for $2\leq m\leq M$.
Set $\bar{\d}:=[\bar{\d}^{(1)\top},\cdots,\bar{\d}^{(M)\top}]^{\top}$, then it follows from (\ref{E:single_basic}) that
\begin{equation}\label{E:single_basic_unknown}
\tilde{\Y}\bar{\mathbf{\Theta}}\bar{\d} = \mathbf{0}.
\end{equation}
We can directly recover $\bar\d$ from \eqref{E:single_basic_unknown} while the identifiability conditions stated in Section \ref{SS:single_known} become harder to satisfy due to overshooting the filter orders. Hence, we recover $\bar\d$ via solving the following problem that is in the same form as \eqref{E:l1_analysis_problem}
\begin{equation}\label{E:l1_analysis_problem_single}
\hat{\bar{\d}} = \argmin_{\bar{\d}} \| \bbDelta \bar{\d} \|_1 \quad \text{s.t. } \tilde{\mathbf{Y}} \bar{\bbTheta} \bar{\d} = \mathbf{0}, \,\, \bar{d}_1 = 1,
\end{equation}
where the diagonal matrix $\mathbf{\Delta}$ contains positive predefined weights, and the second constraint is added to avoid the trivial zero solution.
Notice that the selection of $\mathbf{\Delta}$ should adopt different strategies from \eqref{E:l1_analysis_problem} since $\bar{\d}^{(m)}$ and $\bar{\h}^{(m)}$ have different sparsity patterns.
Furthermore, let us denote by $\mathbf{\Gamma}$ the matrix obtained by dropping the first column of $\tilde{\Y}\bar{\mathbf{\Theta}}$, by $\bar{\d}_{(-1)}$ the vector obtained by dropping the first entry of $\bar{\d}$, and by $\mathbf{\Delta}_{(-1)}$ the matrix obtained by dropping the first column and the first row of $\mathbf{\Delta}$. Define $\mathcal{J}=\mathrm{supp}(\bar{\d}_{(-1)})\subset\{1,\cdots,\sum_{m=1}^M Q_m -1\}$ and $\mathcal{J}^c$ as its complement. Similar to Theorem~\ref{Th:multi_unknown_sufficient}, we can have the following theorem that states the sufficient conditions under which the solution to \eqref{E:l1_analysis_problem_single} is unique and coincides with the true $\bar{\d}$.
\begin{mytheorem}\label{Th:single_unknown_sufficient}
	The solution $\hat{\bar\d}$ to \eqref{E:l1_analysis_problem_single} coincides with the true $\bar{\d}$ if the following two conditions hold:\\
	i) $\rank(\mathbf{\Gamma}_{\mathcal{J}} )= |\mathcal{J}|$, and\\
	ii) There exists a constant $\delta>0$ such that 
		$$\xi = \|\mathbf{I}_{\mathcal{J}^c}^\top(\delta^{-2}\bbDelta_{(-1)}^{-1}\mathbf{\Gamma}^\top\mathbf{\Gamma}\bbDelta_{(-1)}^{-1}+\mathbf{I}_{\mathcal{J}^c}\mathbf{I}_{\mathcal{J}^c}^\top)^{-1}\mathbf{I}_{\mathcal{J}}\|_{\infty} <1.$$
\end{mytheorem}

Similar to Section \ref{SS:multi_unknown}, we can also develop robust recovery guarantees as stated in Theorem \ref{Th:multi_unknown_noise} for problem \eqref{E:l1_analysis_problem_single}, which are omitted here for brevity.

\section{Numerical Experiments}\label{S:num_exp}

The goal of this section is to run numerical experiments to verify the theoretical results, evaluate the performance of the novel schemes for a range of scenarios, and provide insights on their performance. Unless otherwise stated, we consider undirected and connected graphs, and use their adjacency matrices as graph shift operators.\footnote{The code needed to replicate all the numerical experiments here presented can be found at~\url{https://github.com/yuzhu2019/blind_id}.}

\subsection{Multiple Network Processes of Known Filter Orders}

In this section, we estimate multiple network processes with known filter orders and, hence, we adopt the schemes proposed in Section~\ref{SS:multi_known}.

\vspace{1mm}
\noindent \emph{Verification of Theorems~\ref{Th:multi_sufficient}-\ref{Th:multi_sufficient_necessary}.}
We first consider the noiseless case and verify the proposed Theorems~\ref{Th:multi_sufficient}-\ref{Th:multi_sufficient_necessary} via experiments. 
Combining Theorems~\ref{Th:multi_sufficient}, \ref{Th:multi_necessary} and our discussion after Theorem~\ref{Th:multi_sufficient_necessary}, one can conclude that when the polynomials defined by filter coefficients do not share any common roots, the graph filters are identifiable if $|\Omega_1|\geq L_{\max}+L_{\min}-1$ and unidentifiable if $|\Omega_1|<\max\{L_{\max},(\sum_{m=1}^M L_m-1)/(M-1)\}$.  

We consider unweighted Erd\H{o}s-R\'{e}nyi (ER) random graphs \cite{random_graphs} with $N=25$ nodes and edge-formation probability $p=0.2$, and $M=5$ graph filters of identical order $L=8$.
We generate filter coefficients as $\mathbf{h}^{(m)}=a\mathbf{h}^{(1)}+(1-a)\bbn^{(m)}$ for $1<m\leq M$, where the entries in $\mathbf{h}^{(1)}$ and $\bbn^{(m)}$ are independently drawn from the standard normal distribution (SND).
We set $a \in \{0, 0.5, 0.8, 0.9\}$ to impose increasing levels of correlation among filters.
In particular, the coefficients of different filters are uncorrelated when $a=0$. The input $\mathbf{x}=\mathbf{V}\tilde{\mathbf{x}}$ is generated as a bandlimited graph signal, so that the first $|\Omega_1|$ entries of $\tilde{\mathbf{x}}$ are independently drawn from the SND and the remaining ones are set to zero.

For this scenario, only a scaled version of the true graph filters can be recovered. 
Hence, we compute the recovery error as $\|\hat{\mathbf{h}}-\mathbf{h}\|_2/\|\mathbf{h}\|_2$ after setting $\|\hat{\mathbf{h}}\|_2=\|\mathbf{h}\|_2$ and $\hat{\mathbf{h}}^{\top}\mathbf{h}>0$, where $\mathbf{h}$ is the ground truth and $\hat{\mathbf{h}}$ is the estimate.
We consider the recovery to be successful if the recovery error is less than $0.01$.
We consider $|\Omega_1| \in \{1,2,\cdots,24\}$.
For each value of $|\Omega_1|$, we run $1000$ experiments and compute the average recovery rate; see Fig.~\ref{Fig:verify_thm123}.
According to our analysis above, the recovery rate should be one if $|\Omega_1|\geq 15$ and should be zero if $|\Omega_1|\leq 9$.
We can see that the experiment results are consistent with our theoretical results.
Moreover, there is a transition region when $9<|\Omega_1|<15$, and it can be observed that the recovery is more challenging for larger values of $a$.
As expected, when the correlation between graph filters is larger, their identification is more difficult.


\begin{figure*}
	\centering
	\subfigure[]{
		\centering
		\includegraphics[width=0.48\textwidth]{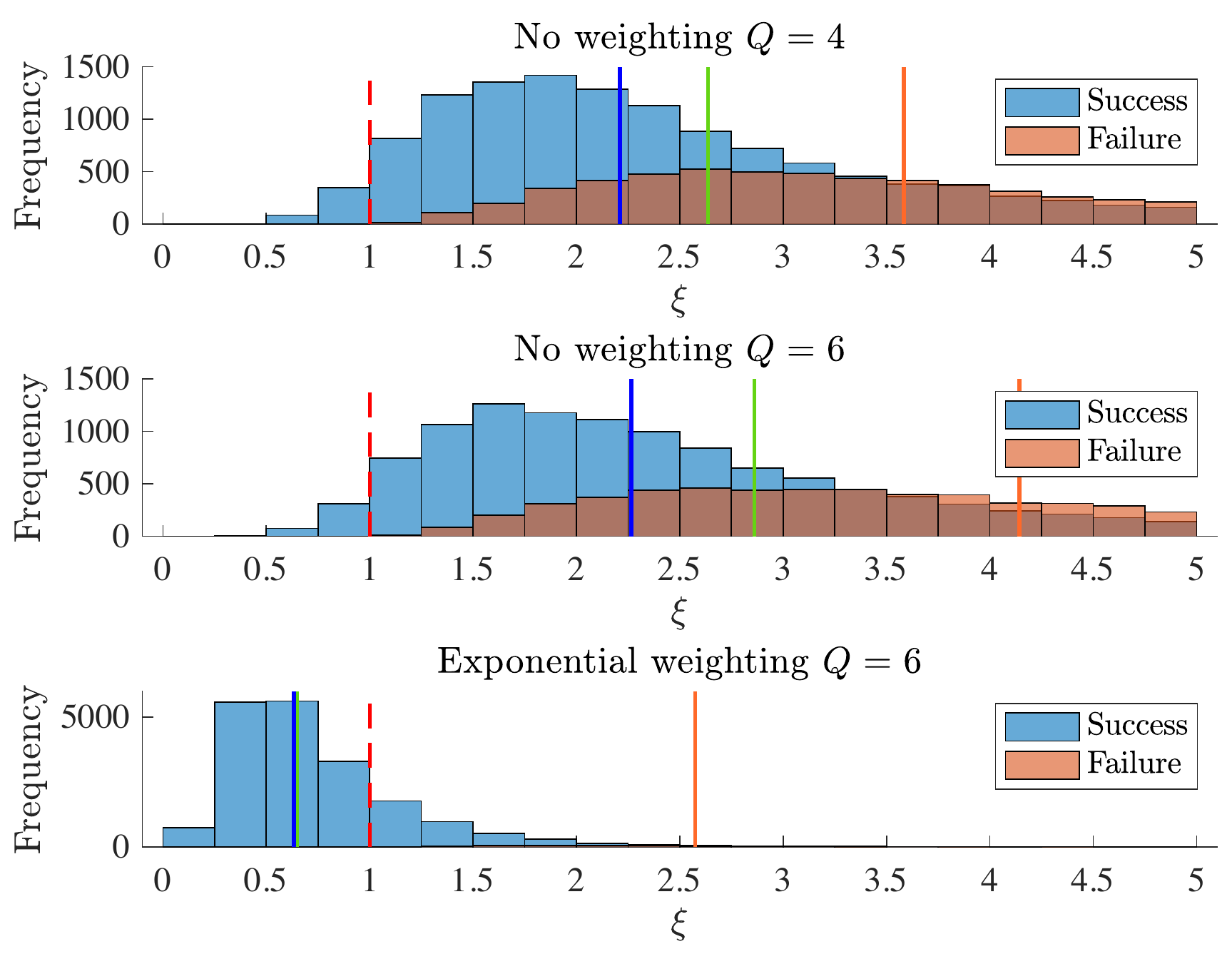}
		\label{Fig:verify_thm4}
	}	
	\subfigure[]{
		\centering
		\includegraphics[width=0.48\textwidth]{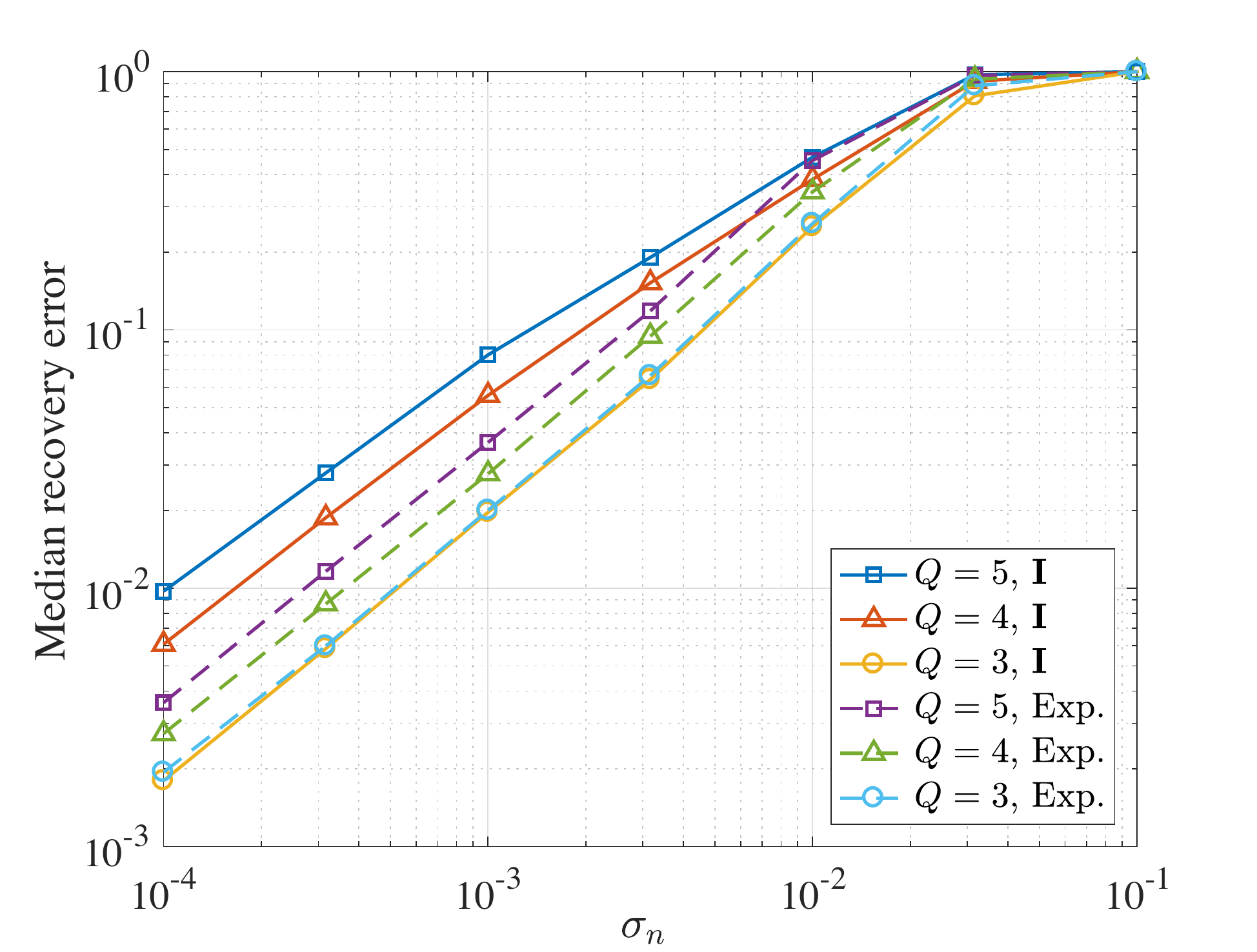}
		\label{Fig:multi_Q_W}
	}
	\vspace{-2mm}
	\caption{\small Estimation of multiple network processes ($M=3$ filters of order $L=3$) when their filter orders are unknown. (a) Experimental validation of Theorem~\ref{Th:multi_unknown_sufficient} together with comparisons of different overshoot filter lengths $Q$ and weight matrices. The height of the bars represents the empirical frequency of $\xi$. A successful recovery is guaranteed whenever $\xi<1$. Blue, orange and green vertical lines show the median values of $\xi$ across successful, failed and all recoveries, respectively. The ratio of successful recoveries for the settings from top to bottom are $0.64$, $0.58$, and $0.97$. A larger $Q$ increases the recovery difficulty while the exponential weights enhance the recovery performance. (b) Median recovery error as a function of the noise level $\sigma$ for different $Q$ and weight matrices.}
	\vspace{-2mm}
\end{figure*}

\vspace{1mm}
\noindent \emph{Influence of the input spectral richness.}
We study the impact of the cardinality of $\Omega_1$ -- which characterizes the spectral richness of the input -- on the recovery error under different noise levels.
{For this, we consider two types of unweighted graphs having the same size $N=30$ and expected value of number of edges: 1) ER graphs with edge-formation probability $p=4/30$; and ii) small-world graphs~\cite{random_graphs} of mean node degree $4$ and rewiring probability $0.2$ generated using a Watts-Strogatz model.}
We consider $M=3$ graph filters of identical order $L=3$.
The filter coefficients are independently drawn from the SND.
The method of generating the input is the same as that used in generating Fig.~\ref{Fig:verify_thm123}, and we set $|\Omega_1| \in \{12,18,24\}$.
The noisy output for process $m$ is generated as $\mathbf{y}^{(m)}=\bbH^{(m)}\bbx+ \gamma \pmb{\omega}^{(m)}$ where the entries of $\pmb{\omega}^{(m)}$ are independently drawn from the SND and $\gamma = \sigma \sqrt{\|\mathbf{H}^{(m)}\bbx\|_2^2/N}$ controls the expected signal-to-noise ratio.
We consider different noise levels by varying $\sigma$ from $10^{-5}$ to $10^{-1}$.
The results are shown in Fig.~\ref{Fig:influence_omega1}.
As expected, the recovery error decreases as $|\Omega_1|$ increases.
To see why this is the case, notice that for larger values of $|\Omega_1|$ there are more cross relations between filters which can be leveraged [cf.~\eqref{E:multi_pair}]. Thus, larger values of $|\Omega_1|$ can enhance the algorithm robustness when noise exists.
{It can also be observed that the performance difference between different graph types becomes smaller for larger $|\Omega_1|$.}
Indeed, for $|\Omega_1|=12$, we can observe that the random nature of ER graphs is beneficial for recovery, whereas this advantage vanishes for larger values of $|\Omega_1|$.

\vspace{1mm}
\noindent \emph{Influence of the number and lengths of filters.}
We study how the number of filters $M$ and the filter lengths (to facilitate exposition we assume $L_m=L$ for all $m$) affect the recovery error under different noise levels.
We adopt the stochastic block model (SBM)~\cite{random_graphs} to generate unweighted graphs with $N=30$ nodes and two blocks of equal size, where the vertex attachment probabilities across blocks and within blocks are $0.1$ and $0.3$, respectively.
Both the filter coefficients and the entries in the input $\mathbf{x}$ are independently drawn from the SND.
We consider five combinations of $M$ and $L$; see Fig.~\ref{Fig:multi_M_L}.

The plot confirms that the recovery accuracy increases as $L$ decreases and as $M$ increases. To better understand this effect, notice that when $M$ is fixed and $L$ increases, the number of cross relations between filters keeps unchanged while the number of unknown filter coefficients increases, and this increases the recovery difficulty.
On the other hand, when $L$ is fixed and $M$ increases, both the number of cross relations and the number of unknown filter coefficients increase, but under our parameter setting the former one increases faster, so the recovery accuracy is improved.

\subsection{Multiple Network Processes of Unknown Filter Orders}

\begin{figure*}
	\centering
	\subfigure[]{
		\centering
		\includegraphics[width=0.43\textwidth]{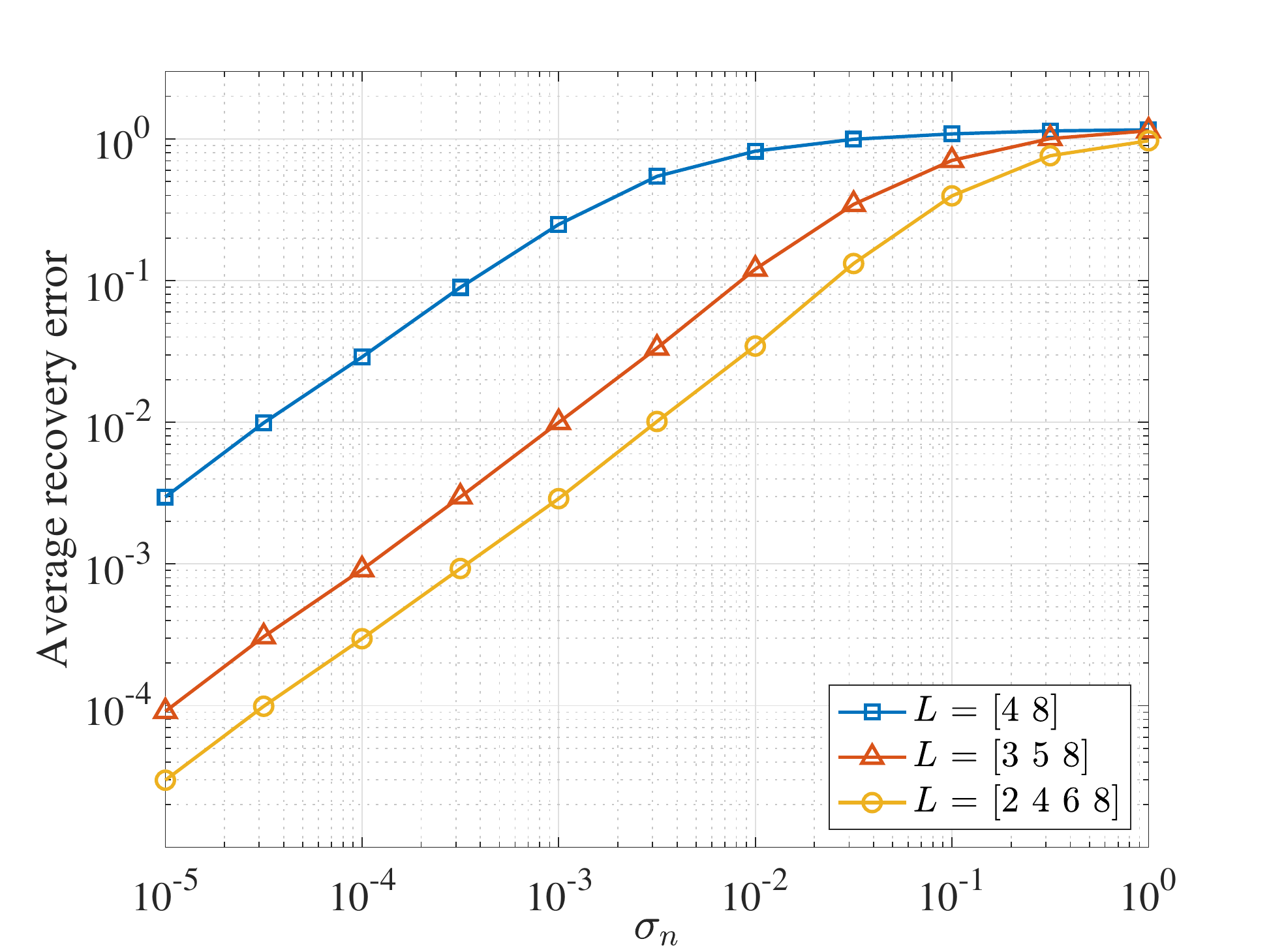}
		\label{Fig:single_orderknown}
	}	
	\hspace{5mm}
	\subfigure[]{
		\centering
		\includegraphics[width=0.43\textwidth]{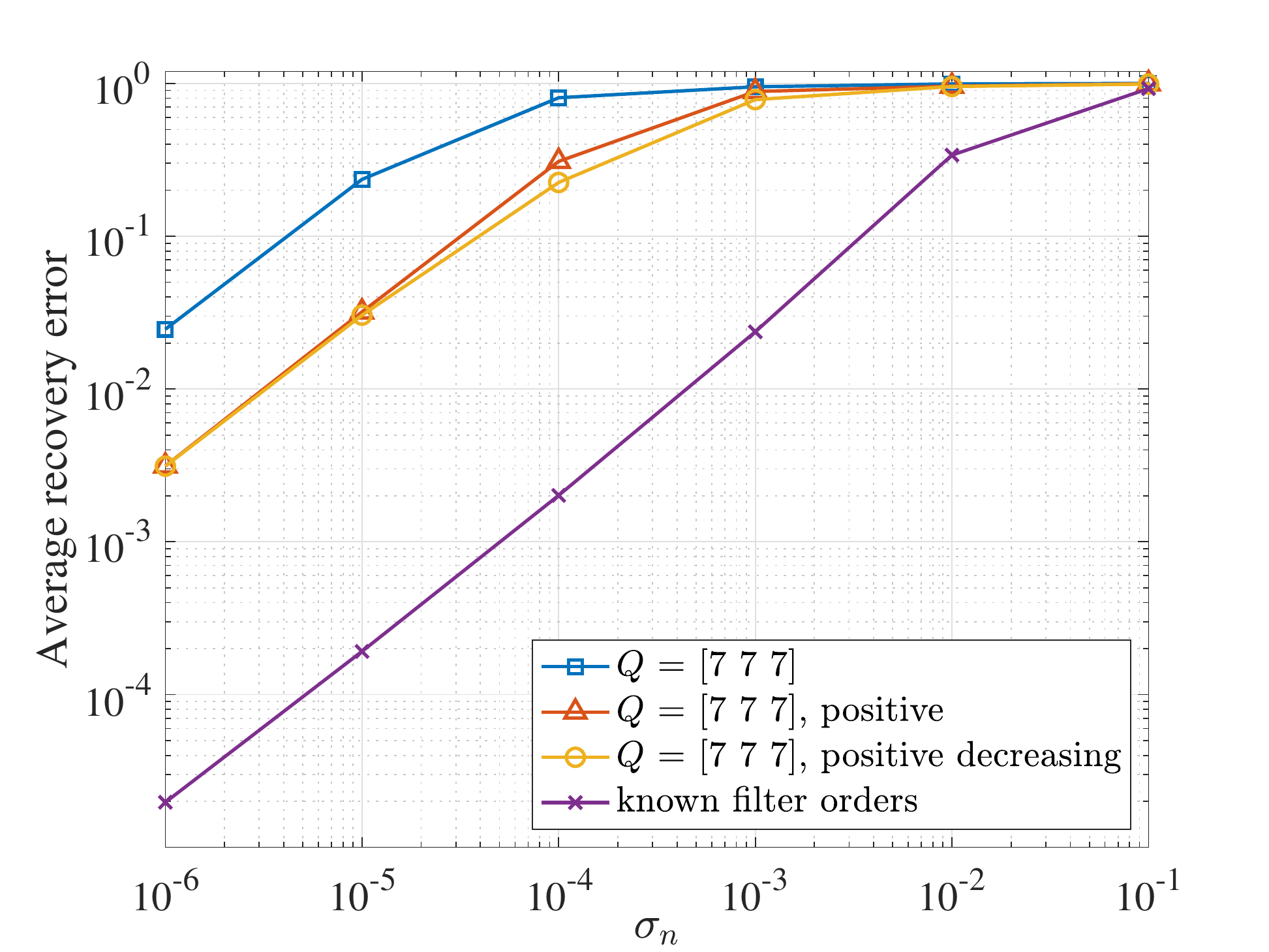}
		\label{Fig:compare}
	}
	\vspace{-2mm}
	\caption{\small Estimation of a single network process. (a) Average recovery error as a function of the noise level $\sigma$ for different numbers of observations when the filter orders are known. When the total filter length is fixed, a larger number of observations results in a higher recovery accuracy. (b) Comparison among algorithms with known or unknown filter orders and extra prior information.}
	\vspace{-2mm}
\end{figure*}

In this section, we implement the method proposed in Section~\ref{SS:multi_unknown} to jointly estimate multiple network processes when their orders are unknown. 
We adopt unweighted ER graphs of size $N=30$ and edge-formation probability $p=0.1$.
We consider $M=3$ graph filters of identical order $L=3$. 
The first entry of the filter $\mathbf{h}^{(1)}$ is set to $1$, and all of the other filter coefficients, as well as the entries in the input $\mathbf{x}$, are independently drawn from the SND.
We consider an identical overshoot filter length $Q$ across all filters.
For the weight matrix $\mathbf{\Delta}$ in \eqref{E:l1_analysis_problem}, we consider two settings: i) no weights, i.e. $\mathbf{\Delta}=\mathbf{I}$, and ii) exponential weights, in which we set $\mathbf{\Delta}_{ii}=\exp(i\,\,\mathrm{mod}\,\,Q)$ and $\mathrm{mod}$ is the modulo operation.

\vspace{1mm}
\noindent \emph{Verification of Theorem~\ref{Th:multi_unknown_sufficient}.}
We first verify Theorem~\ref{Th:multi_unknown_sufficient} in the noiseless case.
Three combinations of $Q$ and $\mathbf{\Delta}$ are considered, as shown in Fig.~\ref{Fig:verify_thm4}.
For each setting, we implement $20,\!000$ realizations, and in each realization, we make sure that condition \emph{i)} in Theorem~\ref{Th:multi_unknown_sufficient} is satisfied.
We plot the number of successes and failures in recovering the graph filters as a function of $\xi$ defined in Theorem~\ref{Th:multi_unknown_sufficient} where we set $\delta=0.02$ (the fraction of realizations for which $\xi>5$ is very small and, thus, those realizations are not shown in the figures).
The recovery is considered to be successful if the recovery error, which is computed as $\|\hat{\bar{\mathbf{h}}}_{(-1)}-\bar{\mathbf{h}}_{(-1)}\|_2/\|\bar{\mathbf{h}}_{(-1)}\|_2$ [cf. \eqref{E:l1_analysis_problem2}], is less than $0.01$.

We can see that, as expected, for all realizations satisfying $\xi<1$ a successful recovery is achieved, and failed recoveries start to appear when $\xi\geq 1$.
Moreover, a smaller value of $\xi$ should imply an easier recovery.
We plot the median values of $\xi$ across successful, failed, and all recoveries using blue, orange and green vertical lines, respectively.
Comparing the two top plots in Fig.~\ref{Fig:verify_thm4}, we find that selecting a larger $Q$ penalizes the recovery due to the increased degrees of freedom.
In addition, the comparison of the bottom two plots in Fig.~\ref{Fig:verify_thm4} shows that the exponential weighting scheme improves the algorithm performance significantly.

\vspace{1mm}
\noindent \emph{Influence of overshoot filter lengths and weight matrices.}
We next explore how different overshoot filter lengths and weight matrices affect the recovery accuracy under different noise levels.
We consider six combinations of $Q$ and $\mathbf{\Delta}$ as shown in Fig.~\ref{Fig:multi_Q_W}, including the case when $Q$ equals the true filter length $L=3$.
We use CVX~\cite{cvx} to solve \eqref{E:multi_unknown_noise} and, recalling the decomposition $\tilde{\Y}\mathbf{\Theta}=[\b, \mathbf{\Phi}]$, we set $\epsilon=\|\mathbf{\Phi}\bar{\mathbf{h}}_{(-1)}+\mathbf{b}\|_2$ where $\bar{\mathbf{h}}_{(-1)}$ is the ground truth.

The results are consistent with those shown in Fig.~\ref{Fig:verify_thm4}.
It can be observed that the recovery accuracy decreases as $Q$ increases due to the increased degrees of freedom. 
For a fixed $Q>L$, selecting an exponential weighting scheme achieves a better recovery than setting $\mathbf{\Delta}=\mathbf{I}$, since the former facilitates the identification of the zeros artificially added.
Moreover, when $Q=L$, the two weighting choices achieve almost the same performance.

\subsection{Single Network Process}

\vspace{1mm}
\noindent \emph{Influence of the number of observations.}
We use the method proposed in Section~\ref{SS:single_known} to estimate a single network process when the filter orders are known and analyze  the influence of the number of observations.
We compare three settings: i)~$L_1=4$, $L_2=8$ (represented as $L=[4\,\,8]$ for simplicity), ii)~$L=[3\,\,5\,\,8]$, and iii)~$L=[2\,\,4\,\,6\,\,8]$.
These settings have the same number of unknown filter coefficients to be estimated but have an increasing number of observations.
We adopt weighted ER graphs of size $N=30$ and edge-formation probability $p=0.1$.
The edge weights are randomly selected from the uniform distribution $\mathrm{U}(0.1,0.7)$.
The filter coefficients and the entries in the input $\mathbf{x}$ are independently drawn from the SND.
The noisy output is generated in the same way as the one used in Fig.~\ref{Fig:influence_omega1}.
The results are shown in Fig.~\ref{Fig:single_orderknown}.
As expected, the recovery accuracy increases as the number of observations increases.

\vspace{1mm}
\noindent \emph{A real-world network.} Lastly, we consider a more practical case. 
We consider a real-world social network, namely the Zachary's karate club network~\cite{zachary1977information}, which contains $N=34$ nodes and $78$ edges.
We assume that there is one information diffusion process on this network.
We generate positive and decreasing filter coefficients modeling the fact that people's opinions are more likely to be affected by the neighbors that are closer to them.
The outputs are observed at $L_1=3$, $L_2=5$ and $L_3=7$, implying that the total filter length is $7$.
The filter coefficients (except for the first one, which is set to $1$) are randomly selected from $\mathrm{U}(0.2,1)$ and then sorted in decreasing order.

We compare four methods: i) assume unknown filter orders and adopt the algorithm proposed in Section~\ref{SS:single_unknown} with $Q_1=Q_2=Q_3=7$; ii) based on method i), add the constraint that the filter coefficients are non-negative to the optimization problem \eqref{E:l1_analysis_problem_single}; iii) based on method i), add the constraints that the filter coefficients are non-negative and decreasing; and iv) assume known filter orders and use the approach proposed in Section~\ref{SS:single_known}.
We plot the average recovery error as a function of the noise level $\sigma$; see Fig.~\ref{Fig:compare}.
It can be observed that, for all methods, the recovery error decreases as $\sigma$ decreases.
The extra prior information, especially the one that the filter coefficients are all positive, helps to improve the recovery accuracy.
As expected, the method leveraging the true filter orders performs best.

\section{Conclusions}\label{S:conclusions}

We investigated the problem of blind identification of multiple graph filters.
This is a generalization of the classical blind multi-channel identification problem to signals defined on graphs and can be of interest in estimating network diffusion processes.
These multiple network processes can indeed correspond to different processes defined on the same network or a single process that is sensed at different points in time.
For both scenarios, we considered two cases where the filter orders are known and unknown, respectively.
A least-squares approach was advocated for the former case and a sparse recovery algorithm was proposed for the latter case. 
Recovery conditions and theoretical guarantees were also provided.
Numerical experiments demonstrated the effectiveness of the proposed methods and validated our theoretical claims.
Current and future research avenues include: i) the estimation of multiple network processes defined on different (but related) graphs; ii) the joint estimation of the network topology and the specifications of the network processes when the underlying graph is (partially) unknown; iii) the identification of non-linear network processes; and iv) the application of the developed techniques to real-world datasets, especially in the field of neuroscience.

\begin{appendices}

\section{Proof of Theorem~\ref{Th:multi_sufficient} (Cont.)}\label{A:repeated_roots}

We are left with the task of showing that $\hat{p}^{(m)}(z) = \alpha {p}^{(m)}(z)$ for the case where ${p}^{(m)}(z)$ has repeated roots.
Assuming that $p^{(m)}(z)$ has a repeated root $z_0$ of multiplicity $k>1$, it follows that the $i$th order derivative of $p^{(m)}(z)\hat{p}^{(n)}(z)$ equals zero for $i=0,\cdots,k-1$, i.e.,
\begin{equation}\label{E:repeat_root_1}
\left.\frac{d^i \left(p^{(m)}(z)\hat{p}^{(n)}(z)\right)}{dz^i}\right|_{z=z_0}=0.
\end{equation}
From (\ref{E:multi_known_suffi_proof_020}), we have 
\begin{equation}\label{E:repeat_root_2}
\frac{d^i \left( p^{(m)}(z)\hat{p}^{(n)}(z) - p^{(n)}(z)\hat{p}^{(m)}(z) \right)}{dz^i} = 0
\end{equation}
for all $z\in\mathbb{C}$ and thus it also holds for $z=z_0$. Combining (\ref{E:repeat_root_1}) and (\ref{E:repeat_root_2}), we have 
\begin{equation}\label{E:repeat_root_3}
\left.\frac{d^i \left(p^{(n)}(z)\hat{p}^{(m)}(z)\right)}{dz^i}\right|_{z=z_0}=0.
\end{equation}
When $i=1$, \eqref{E:repeat_root_3} can be rewritten as
\begin{equation}
\left.\frac{dp^{(n)}(z)}{dz} \hat{p}^{(m)}(z) + p^{(n)}(z)\frac{d\hat{p}^{(m)}(z)}{dz} \right|_{z=z_0}= 0,
\end{equation}
from where it follows that $\left.\frac{d\hat{p}^{(m)}(z)}{dz} \right|_{z=z_0}=0$ since $\hat{p}^{(m)}(z_0)=0$ and there must exist one $n^{\ast}$ such that $p^{(n^{\ast})}(z_0)\neq 0$. 
When $i=2$, \eqref{E:repeat_root_3} can be rewritten as
\begin{align}
	\frac{d^2p^{(n)}(z)}{dz^2}& \hat{p}^{(m)}(z) + 2\frac{dp^{(n)}(z)}{dz}\frac{d\hat{p}^{(m)}(z)}{dz} \nonumber\\
						    &	\left. + p^{(n)}(z)\frac{d^2\hat{p}^{(m)}(z)}{dz^2}\right|_{z=z_0}= 0,
\end{align}
and it follows that $\frac{d^2\hat{p}^{(m)}(z)}{dz^2}=0$ since $\hat{p}^{(m)}(z_0)=0$, $\left.\frac{d\hat{p}^{(m)}(z)}{dz} \right|_{z=z_0}=0$ and there must exist one $n^{\ast}$ such that $p^{(n^{\ast})}(z_0)\neq 0$.
Similarly, we can prove that $\left.\frac{d^i\hat{p}^{(m)}(z)}{dz^i} \right|_{z=z_0}=0$ for $i=3,\cdots,k-1$ successively by leveraging (\ref{E:repeat_root_3}) and the results obtained in the previous $i-1$ steps. 
This means that $z_0$ is also a repeated root of $\hat{p}^{(m)}(z)$ of multiplicity $k$. Hence, for the case of repeated roots, we still have that $p^{(m)}(z)$ and $\hat{p}^{(m)}(z)$ share the same roots (and multiplicity), from where it follows that $\hat{\h}=\alpha\h$.

\section{Proof of Theorem \ref{Th:multi_sufficient_necessary}}\label{A:proof_multi_known}

The proof is inspired by the proof of \cite[Thm. 3]{xu_1995_least}.

\vspace{1mm}
\noindent \emph{Necessary Part:} 
We prove this part by showing that if the condition stated in Theorem \ref{Th:multi_sufficient_necessary} does not hold, the solution to (\ref{E:multi_basic}) is not identifiable.
Clearly, the vector $[1,\cdots,1,-\h^{\top}]^{\top}$ of size $|\Omega_1|+\sum_{m=1}^M L_m$ is in the null space of the matrix defined in (\ref{E:multi_matrix}).
If the condition stated in Theorem \ref{Th:multi_sufficient_necessary} does not hold, i.e., there exists another independent vector $[g_1,\cdots,g_{|\Omega_1|},-\hat{\h}^{\top}]^{\top}$ which is also in the null space of the matrix in (\ref{E:multi_matrix}), then it follows that $\hat\p(z_i)=g_i\p(z_i)$ and $\hat{p}^{(m)}(z_i)=g_i p^{(m)}(z_i)$ for $1\leq i\leq |\Omega_1|$ and $1\leq m\leq M$. 
Notice that $\hat\p(\cdot)$ and $\hat{p}^{(m)}(\cdot)$ are defined in a similar way to $\p(\cdot)$ and ${p}^{(m)}(\cdot)$ but associated with $\hat\h$. 
Then, for any pair of two filters $m$ and $n$, we have that
\begin{small}
	\begin{equation}\label{E:proof_multi_known_1}
		\hat{p}^{(m)}(z_i)p^{(n)}(z_i)= g_i p^{(m)}(z_i)p^{(n)}(z_i)=\hat{p}^{(n)}(z_i)p^{(m)}(z_i).
	\end{equation}
\end{small}
Hence, $\hat\h$ is another solution.

We are left to show that the solutions $\hat{\h}$ and $\h$ are indeed not linearly dependent, i.e., that $\hat{\h} \neq \alpha \h$ for every scalar $\alpha$. To show this, we assume that $\hat{\h} = \alpha \h$ and arrive to a contradiction. From our assumption it follows that $(\alpha-g_i)\p(z_i)=\mathbf{0}$, which implies that $\alpha-g_i=0$ or $\p(z_i)=\mathbf{0}$. If $\alpha-g_i=0$ for all $1\leq i\leq |\Omega_1|$, the two vectors $[1,\cdots,1,-\h^{\top}]^{\top}$ and $[g_1,\cdots,g_{|\Omega_1|},-\hat{\h}^{\top}]^{\top}$ are dependent, which contradicts our assumption. Hence, there exits at least one $j$ satisfying that $g_j\neq \alpha$ and $\p(z_j)=\mathbf{0}$. This means that all filters share one common root $z_j$, and in this case, the solution is not identifiable according to Theorem \ref{Th:multi_necessary}.

\vspace{1mm}
\noindent \emph{Sufficient Part:}
We first show that if the condition stated in Theorem \ref{Th:multi_sufficient_necessary} holds, there does not exist a root shared by all the filters.
Suppose that there exists a root $z_0$ shared by all the filters, we can write $p^{(m)}(z)=q^{(m)}(z)(z-z_0)$.
If $z_0\in\pmb{\lambda}_{\Omega_1}$, the matrix defined in (\ref{E:multi_matrix}) will have a column whose entries are all zero, then the condition stated in Theorem \ref{Th:multi_sufficient_necessary} does not hold. 
If $z_0\notin\pmb{\lambda}_{\Omega_1}$, we set $\hat{p}^{(m)}(z)=q^{(m)}(z)(z-z'_0)$ where $z'_0\neq z_0$ and $1\leq m\leq M$.
Then it follows that 
	\begin{equation}\label{E:proof_multi_known_2}
		\hat{p}^{(m)}(z_i)=q^{(m)}(z_i)(z_i-z_0)\frac{z_i-z'_0}{z_i-z_0} =  g_ip^{(m)}(z_i),
	\end{equation}
where we set $g_i=\frac{z_i-z'_0}{z_i-z_0}$. Then, $[g_1,\cdots,g_{|\Omega_1|},-\hat{\h}^{\top}]^{\top}$ will be another independent null vector of the matrix defined in (\ref{E:multi_matrix}), which contradicts the condition stated in Theorem \ref{Th:multi_sufficient_necessary}.

Next, we show that if the condition stated in Theorem \ref{Th:multi_sufficient_necessary} holds, the solution is identifiable. 
We show this by proving its contrapositive statement, i.e., if the solution is not identifiable then the matrix in \eqref{E:multi_matrix} does not have a one-dimensional null space.
If the solution cannot be uniquely identified, there exits another solution $\hat\h$ independent of $\h$ and satisfying (\ref{E:multi_poly}). Then we have that
	\begin{equation}\label{E:proof_multi_known_3}
		\underbrace{\left[\begin{matrix}
		\P_1 & & \mathbf{0} \\
		& \ddots & \\
		\mathbf{0} & & \P_{|\Omega_1|}
		\end{matrix}\right]}_{\check{\P}}
		\underbrace{\left[\begin{matrix}
		\Z(z_1)\\
		\vdots \\
		\Z(z_{|\Omega_1|})
		\end{matrix}\right]}_{\check{\Z}}
		\hat\h=\mathbf{0},
	\end{equation}
where $\P_i$ of dimension $(M-1)\times M$ is defined as
	{\small $$
	\left[\begin{matrix}
	-p^{(l_i)}(z_i) \!\!&\!\!  \!\!&\!\!  \!\!&\!\! p^{(1)}(z_i) \!\!&\!\!  \!\!&\!\!  \!\!&\! \!\\
	\!\!&\!\! \ddots \!\!&\!\!  \!\!&\!\! \vdots \!\!&\!\!  \!\!&\!\!  \!\!&\!\!\\
	\!\!&\!\!  \!\!&\!\! -p^{(l_i)}(z_i) \!\!&\!\! p^{(l_i-1)}(z_i)\!\!&\!\!  \!\!&\!\!  \!\!&\! \!\\
	\!\!&\!\!  \!\!&\!\! \!\!&\!\! -p^{(l_i+1)}(z_i) \!\!&\!\! p^{(l_i)}(z_i)\!\!&\!\! \!\!&\!\!\\
	\!\!&\!\! \!\!&\!\! \!\!&\!\!\vdots \!\!&\!\! \!\!&\!\!\ddots\!\!&\!\!\\
	\!\!&\!\! \!\!&\!\!  \!\!&\!\!-p^{(M)}(z_i) \!\!&\!\! \!\!&\!\! \!\!&\!\! p^{(l_i)}(z_i)
	\end{matrix}\right]
	$$}where $l_i\in\{1,\cdots,M\}$ and satisfies $p^{(l_i)}(z_i)\neq 0$. We can always find such $l_i$, otherwise $z_i$ will be a common root shared by all filters.

It can be observed that $\rank(\P_i)=M-1$ and $\P_i$ has a single null vector $\p(z_i)$ up to some scalar multiple. 
Hence, the null space of $\check{\P}$ is the column space of the following matrix
	\begin{equation}\label{E:proof_multi_known_4}
		\left[\begin{matrix}
		\p(z_1) & & \\
		&\ddots &\\
		&& \p(z_{|\Omega_1|})
		\end{matrix}\right].
	\end{equation}
From (\ref{E:proof_multi_known_3}), it follows that $\check{\Z}\hat\h$ should be in the null space of $\check{\P}$, which implies that there exist scalars $g_1, \ldots, g_{|\Omega_1|}$ such that
	\begin{equation}\label{E:proof_multi_known_5}
		\left[\begin{matrix}
		\p(z_1) & & &\Z(z_1)\\
		&\ddots & &\vdots \\
		&& \p(z_{|\Omega_1|})&\Z(z_{|\Omega_1|})
		\end{matrix}\right]
		\left[\begin{matrix}
		g_1 \\
		\vdots\\
		g_{|\Omega_1|}\\
		-\hat\h
		\end{matrix}\right]=\mathbf{0}.
	\end{equation}
Since $[1,\cdots,1,-\h^{\top}]^{\top}$ is another independent null vector, the matrix defined in (\ref{E:multi_matrix}) has at least a two-dimensional null space, which contradicts the condition stated in Theorem \ref{Th:multi_sufficient_necessary}.

\section{Proof of Theorem \ref{Th:multi_unknown_sufficient}}\label{A:proof_multi_unknown}

We show that the conditions in the theorem guarantee that the solution to \eqref{E:l1_analysis_problem2} coincides with the true ${\bar{\bbh}}_{(-1)}$. Due to the equivalence between problems \eqref{E:l1_analysis_problem} and \eqref{E:l1_analysis_problem2}, the result follows.

The following two conditions are required for the solution of \eqref{E:l1_analysis_problem2} to coincide with the true filter coefficients $\bar{\bbh}_{(-1)}$ \cite[Thm. 1]{zhang_2016_one}:\\
a) $\mathrm{Ker}( (\bbDelta_{(-1)})_{\mathcal{I}^c}^{\top} )\cap\mathrm{Ker}(\bbPhi)=\{\mathbf{0}\}$, and\\
b) There exists $\bbz\in\mathbb{R}^{\sum_{m=1}^M Q_m-1}$ such that $\bbDelta_{(-1)}\bbz\in\mathrm{Im}(\bbPhi^{\top})$, $\bbz_{\mathcal{I}}=\mathrm{sign}((\bbDelta_{(-1)})_{\mathcal{I}}^{\top}\bar{\bbh}_{(-1)})$, and $\|\bbz_{\mathcal{I}^c}\|_{\infty}<1$,\\
where $\mathrm{Ker}(\cdot)$ and $\mathrm{Im}(\cdot)$ denote the null and column spaces of the argument matrix respectively. The proof now reduces to showing that conditions \emph{i)} and \emph{ii)} in Theorem \ref{Th:multi_unknown_sufficient} imply conditions a) and b) above.

First, we show that \emph{i)} in Theorem \ref{Th:multi_unknown_sufficient} implies a). Since $\bbDelta_{(-1)}$ is a diagonal matrix with positive diagonal entries, $\mathrm{Ker}( (\bbDelta_{(-1)})_{\mathcal{I}^c}^{\top} )$ is spanned by the columns of $\bbI_{\mathcal{I}}$. For a) to hold, we need the $|\mathcal{I}|$ columns of $\bbPhi$ in positions $\mathcal{I}$ to form a full column rank matrix, i.e. condition \emph{i)} in Theorem \ref{Th:multi_unknown_sufficient}. 

Next, we will show that \emph{ii)} in Theorem \ref{Th:multi_unknown_sufficient} implies b). We construct the following $\ell_2$-norm minimization problem and we will show later that its solution satisfies b) if \emph{ii)} in Theorem \ref{Th:multi_unknown_sufficient} holds:
\begin{equation}\label{E:dual_problem}
\min_{\{\bbz,\bbu\}} \delta^2\|\bbu\|_2^2+\|\bbz\|_2^2 \quad\text{s.t. } \bbz = \bbDelta_{(-1)}^{-1}\bbPhi^{\top}\bbu,\,\, \bbz_{\mathcal{I}}=\mathrm{sign}((\bar{\bbh}_{(-1)})_{\mathcal{I}}),
\end{equation}
where $\delta$ is a positive tuning constant. The two constraints in problem (\ref{E:dual_problem}) guarantee the first two requirements in b), hence we are left to show that $\|\bbz_{\mathcal{I}^c}\|_{\infty}<1$.
From the first constraint in (\ref{E:dual_problem}), we have that
\begin{equation}\label{E:constraint1}
\bbDelta_{(-1)}^{-1}\bbPhi^{\top}\bbu = \bbI_{\mathcal{I}}\bbz_{\mathcal{I}} + \bbI_{\mathcal{I}^c}\bbz_{\mathcal{I}^c} =\bbI_{\mathcal{I}}\mathrm{sign}((\bar{\bbh}_{(-1)})_{\mathcal{I}}) + \bbI_{\mathcal{I}^c}\bbz_{\mathcal{I}^c}.
\end{equation}
Defining $\pmb{\Upsilon}=[\delta^{-1}\bbDelta_{(-1)}^{-1}\bbPhi^{\top},\bbI_{\mathcal{I}^c}]$ and $\bbt=[\delta\bbu^{\top},-\bbz_{\mathcal{I}^c}^{\top}]^{\top}$, we get
\begin{equation}
\bbI_{\mathcal{I}}\mathrm{sign}((\bar{\bbh}_{(-1)})_{\mathcal{I}}) = \bbDelta_{(-1)}^{-1}\bbPhi^{\top}\delta^{-1}\delta\bbu - \bbI_{\mathcal{I}^c}\bbz_{\mathcal{I}^c}=\pmb{\Upsilon}\bbt.
\end{equation}
Then we can rewrite (\ref{E:dual_problem}) as 
\begin{equation}\label{E:dual_problem_rewrite}
\min_{\bbt} \|\bbt\|_2^2 \quad\text{s.t. } \pmb{\Upsilon}\bbt = \bbI_{\mathcal{I}}\mathrm{sign}((\bar{\bbh}_{(-1)})_{\mathcal{I}}).
\end{equation}
The solution to (\ref{E:dual_problem_rewrite}) is given by $\bar{\bbt}=\pmb{\Upsilon}^{\dag}\bbI_{\mathcal{I}}\mathrm{sign}((\bar{\bbh}_{(-1)})_{\mathcal{I}})$, where $\pmb{\Upsilon}^{\dag}=\pmb{\Upsilon}^{\top}(\pmb{\Upsilon}\pmb{\Upsilon}^{\top})^{-1}=\pmb{\Upsilon}^{\top}(\delta^{-2}\bbDelta_{(-1)}^{-1}\bbPhi^{\top}\bbPhi\bbDelta_{(-1)}^{-1}+\bbI_{\mathcal{I}^c}\bbI_{\mathcal{I}^c}^{\top})^{-1}$. Condition a) guarantees the existence of the inverse. Hence, we have that
\begin{equation}
\bbz_{\mathcal{I}^c} = - \bbI_{\mathcal{I}^c}^{\top}(\delta^{-2}\bbDelta_{(-1)}^{-1}\bbPhi^{\top}\bbPhi\bbDelta_{(-1)}^{-1}+\bbI_{\mathcal{I}^c}\bbI_{\mathcal{I}^c}^{\top})^{-1}\bbI_{\mathcal{I}}\mathrm{sign}((\bar{\bbh}_{(-1)})_{\mathcal{I}}).
\end{equation}
Since $\|\mathrm{sign}((\bar{\bbh}_{(-1)})_{\mathcal{I}})\|_{\infty}=1$, we can see that \emph{ii)} in Theorem \ref{Th:multi_unknown_sufficient} enforces $\|\bbz_{\mathcal{I}^c}\|_{\infty}<1$, which completes the proof.

\section{Proof of Theorem \ref{Th:multi_unknown_noise}}\label{A:proof_multi_noise}
Theorem \ref{Th:multi_unknown_noise} can be obtained by applying the second claim in \cite[Thm. 2]{zhang_2016_one}.
Since $\mathbf{\Phi}$ does not meet the full row rank assumption in \cite{zhang_2016_one}, $C_3$ depends on its pseudo-inverse here instead of its inverse in \cite{zhang_2016_one}. 
We also use $\|\mathbf{z}_{\mathcal{I}^c}\|_{\infty}\leq\xi$ and $\|\z\|_2<\sqrt{\sum_{m=1}^M Q_m}$ to make $C_2$ and $C_3$ independent of the dual certificate $\mathbf{z}$; see Appendix \ref{A:proof_multi_unknown}.

\end{appendices}

\bibliographystyle{IEEEtran}
\bibliography{multifilter_ID}

\begin{thebibliography}{10}
\providecommand{\url}[1]{#1}
\csname url@samestyle\endcsname
\providecommand{\newblock}{\relax}
\providecommand{\bibinfo}[2]{#2}
\providecommand{\BIBentrySTDinterwordspacing}{\spaceskip=0pt\relax}
\providecommand{\BIBentryALTinterwordstretchfactor}{4}
\providecommand{\BIBentryALTinterwordspacing}{\spaceskip=\fontdimen2\font plus
\BIBentryALTinterwordstretchfactor\fontdimen3\font minus
  \fontdimen4\font\relax}
\providecommand{\BIBforeignlanguage}[2]{{%
\expandafter\ifx\csname l@#1\endcsname\relax
\typeout{** WARNING: IEEEtran.bst: No hyphenation pattern has been}%
\typeout{** loaded for the language `#1'. Using the pattern for}%
\typeout{** the default language instead.}%
\else
\language=\csname l@#1\endcsname
\fi
#2}}
\providecommand{\BIBdecl}{\relax}
\BIBdecl

\bibitem{zhu_2019_estimation}
Y.~{Zhu}, F.~J. {Iglesias}, A.~G. {Marques}, and S.~{Segarra}, ``Estimation of
  network processes via blind graph multi-filter identification,'' in
  \emph{IEEE Int. Conf. on Acoustics, Speech and Signal Process.}, May 2019,
  pp. 5451--5455.

\bibitem{garas_2010_worldwide}
A.~Garas, P.~Argyrakis, C.~Rozenblat, M.~Tomassini, and S.~Havlin, ``Worldwide
  spreading of economic crisis,'' \emph{New Journal of Physics}, vol.~12,
  no.~11, p. 113043, Nov. 2010.

\bibitem{liben_2007_link}
D.~Liben-Nowell and J.~Kleinberg, ``The link-prediction problem for social
  networks,'' \emph{J. Am. Soc. Info. Sci. Technol.}, vol.~58, no.~7, pp.
  1019--1031, 2007.

\bibitem{kleinberg_1999_authorative}
J.~M. Kleinberg, ``Authoritative sources in a hyperlinked environment,''
  \emph{J. ACM}, vol.~46, no.~5, pp. 604--632, Sep. 1999.

\bibitem{balthrop_2004_technological}
J.~Balthrop, S.~Forrest, M.~E.~J. Newman, and M.~M. Williamson, ``Technological
  networks and the spread of computer viruses,'' \emph{Science}, vol. 304, no.
  5670, pp. 527--529, Apr. 2004.

\bibitem{bu_2003_topological}
D.~Bu, Y.~Zhao, L.~Cai, H.~Xue, X.~Zhu, H.~Lu, J.~Zhang, S.~Sun, L.~Ling,
  N.~Zhang, G.~Li, and R.~Chen, ``Topological structure analysis of the
  protein--protein interaction network in budding yeast,'' \emph{Nucleic Acids
  Res.}, vol.~31, no.~9, pp. 2443--2450, 2003.

\bibitem{medaglia_2017_brain}
J.~D. Medaglia, W.~Huang, S.~Segarra, C.~Olm, J.~Gee, M.~Grossman, A.~Ribeiro,
  C.~T. McMillan, and D.~S. Bassett, ``Brain network efficiency is influenced
  by the pathologic source of corticobasal syndrome,'' \emph{Neurology},
  vol.~89, no.~13, pp. 1373--1381, 2017.

\bibitem{sandryhaila_2013_discrete}
A.~Sandryhaila and J.~M.~F. Moura, ``Discrete signal processing on graphs,''
  \emph{IEEE Trans. Signal Process.}, vol.~61, no.~7, pp. 1644--1656, Apr.
  2013.

\bibitem{shuman_2013_emerging}
D.~I. Shuman, S.~K. Narang, P.~Frossard, A.~Ortega, and P.~Vandergheynst, ``The
  emerging field of signal processing on graphs: Extending high-dimensional
  data analysis to networks and other irregular domains,'' \emph{IEEE Signal
  Process. Mag.}, vol.~30, no.~3, pp. 83--98, 2013.

\bibitem{ortega_2018_graph}
A.~Ortega, P.~Frossard, J.~Kovacevic, J.~M.~F. Moura, and P.~Vandergheynst,
  ``Graph signal processing: Overview, challenges, and applications,''
  \emph{Proc. IEEE}, vol. 106, no.~5, pp. 808--828, May 2018.

\bibitem{marques_2016_sampling}
A.~G. Marques, S.~Segarra, G.~Leus, and A.~Ribeiro, ``Sampling of graph signals
  with successive local aggregations,'' \emph{IEEE Trans. Signal Process.},
  vol.~64, no.~7, pp. 1832--1843, Apr. 2016.

\bibitem{chen_2015_sampling}
S.~Chen, R.~Varma, A.~Sandryhaila, and J.~Kovacevic, ``Discrete signal
  processing on graphs: Sampling theory,'' \emph{IEEE Trans. Signal Process.},
  vol.~63, no.~24, pp. 6510--6523, Dec. 2015.

\bibitem{anis_2016_sampling}
A.~Anis, A.~Gadde, and A.~Ortega, ``Efficient sampling set selection for
  bandlimited graph signals using graph spectral proxies,'' \emph{IEEE Trans.
  Signal Process.}, vol.~64, no.~14, pp. 3775--3789, Jul. 2016.

\bibitem{chamon_2018_sampling}
L.~F.~O. Chamon and A.~Ribeiro, ``Greedy sampling of graph signals,''
  \emph{IEEE Trans. Signal Process.}, vol.~66, no.~1, pp. 34--47, Jan. 2018.

\bibitem{romero_2017_reconstruction}
D.~Romero, M.~Ma, and G.~B. Giannakis, ``Kernel-based reconstruction of graph
  signals,'' \emph{IEEE Trans. Signal Process.}, vol.~65, no.~3, pp. 764--778,
  Feb. 2017.

\bibitem{hammond_2011_wavelet}
D.~K. Hammond, P.~Vandergheynst, and R.~Gribonval, ``Wavelets on graphs via
  spectral graph theory,'' \emph{Applied and Comput. Harmonic Anal.}, vol.~30,
  no.~2, pp. 129--150, 2011.

\bibitem{shafipour_2017_fourier}
R.~Shafipour, A.~Khodabakhsh, G.~Mateos, and E.~Nikolova, ``A digraph {Fourier}
  transform with spread frequency components,'' in \emph{Global Conf. Signal
  and Info. Process. (GlobalSIP)}, Nov. 2017, pp. 583--587.

\bibitem{dong_2016_laplacian}
X.~Dong, D.~Thanou, P.~Frossard, and P.~Vandergheynst, ``Learning {L}aplacian
  matrix in smooth graph signal representations,'' \emph{IEEE Trans. Signal
  Process.}, vol.~64, no.~23, pp. 6160--6173, Aug. 2016.

\bibitem{kalofolias_2016_smooth}
V.~Kalofolias, ``How to learn a graph from smooth signals,'' in \emph{Intl.
  Conf. Artif. Intel. Stat. (AISTATS)}, 2016, pp. 920--929.

\bibitem{segarra_2017_topo}
S.~Segarra, A.~Marques, G.~Mateos, and A.~Ribeiro, ``Network topology inference
  from spectral templates,'' \emph{IEEE Trans. Signal Inf. Process. Netw.},
  vol.~3, no.~3, pp. 467--483, Aug. 2017.

\bibitem{shen_2017_kernel}
Y.~Shen, B.~Baingana, and G.~B. Giannakis, ``Kernel-based structural equation
  models for topology identification of directed networks,'' \emph{IEEE Trans.
  Signal Process.}, vol.~65, no.~10, pp. 2503--2516, Feb. 2017.

\bibitem{mateos_2019_connecting}
G.~{Mateos}, S.~{Segarra}, A.~G. {Marques}, and A.~{Ribeiro}, ``Connecting the
  dots: Identifying network structure via graph signal processing,'' \emph{IEEE
  Signal Process. Mag.}, vol.~36, no.~3, pp. 16--43, May 2019.

\bibitem{segarra_2017_filters}
S.~Segarra, A.~G. Marques, and A.~Ribeiro, ``Optimal graph-filter design and
  applications to distributed linear network operators,'' \emph{IEEE Trans.
  Signal Process.}, vol.~65, no.~15, pp. 4117--4131, Aug. 2017.

\bibitem{isufi_2017_filters}
E.~Isufi, A.~Loukas, A.~Simonetto, and G.~Leus, ``Autoregressive moving average
  graph filtering,'' \emph{IEEE Trans. Signal Process.}, vol.~65, no.~2, pp.
  274--288, Jan. 2017.

\bibitem{teke_2017_filters}
O.~Teke and P.~P. Vaidyanathan, ``Extending classical multirate signal
  processing theory to graphs -- {Part I}: Fundamentals,'' \emph{IEEE Trans.
  Signal Process.}, vol.~65, no.~2, pp. 409--422, Jan. 2017.

\bibitem{narang_2012_filter}
S.~K. Narang and A.~Ortega, ``Perfect reconstruction two-channel wavelet filter
  banks for graph structured data,'' \emph{IEEE Trans. Signal Process.},
  vol.~60, no.~6, pp. 2786--2799, Jun. 2012.

\bibitem{sakiyama_2014_oversampled}
A.~Sakiyama and Y.~Tanaka, ``Oversampled graph laplacian matrix for graph
  filter banks,'' \emph{IEEE Trans. Signal Process.}, vol.~62, no.~24, pp.
  6425--6437, Dec. 2014.

\bibitem{tay_2015_design}
D.~B.~H. Tay and Z.~Lin, ``Design of near orthogonal graph filter banks,''
  \emph{IEEE Signal Process. Lett.}, vol.~22, no.~6, pp. 701--704, Jun. 2015.

\bibitem{segarra_2017_design}
S.~Segarra, A.~G. Marques, G.~R. Arce, and A.~Ribeiro, ``Design of weighted
  median graph filters,'' in \emph{IEEE Intl. Wrksp. Computat. Advances
  Multi-Sensor Adaptive Process. (CAMSAP)}, Dec. 2017, pp. 1--5.

\bibitem{xiao_2018_nonlinear}
Z.~Xiao and X.~Wang, ``Nonlinear polynomial graph filter for signal processing
  with irregular structures,'' \emph{IEEE Trans. Signal Process.}, vol.~66,
  no.~23, pp. 6241--6251, Dec. 2018.

\bibitem{segarra_2016_center}
S.~Segarra, A.~G. Marques, G.~R. Arce, and A.~Ribeiro, ``Center-weighted median
  graph filters,'' in \emph{Global Conf. Signal and Info. Process.
  (GlobalSIP)}, Dec. 2016, pp. 336--340.

\bibitem{segarra_2016_blind}
S.~Segarra, G.~Mateos, A.~G. Marques, and A.~Ribeiro, ``Blind identification of
  graph filters,'' \emph{IEEE Trans. Signal Process.}, vol.~65, no.~5, pp.
  1146--1159, Mar. 2017.

\bibitem{ramirez_2017_graph}
D.~Ram{\'\i}rez, A.~G. Marques, and S.~Segarra, ``Graph-signal reconstruction
  and blind deconvolution for diffused sparse inputs,'' in \emph{IEEE Int.
  Conf. on Acoustics, Speech and Signal Process.}, Mar. 2017, pp. 4104--4108.

\bibitem{mateos_2018_blind}
C.~Ye, R.~Shafipour, and G.~Mateos, ``Blind identification of invertible graph
  filters with multiple sparse inputs,'' \emph{arXiv:1803.04072}, 2018.

\bibitem{iglesias_demixing_2018}
F.~J. Iglesias, S.~Segarra, S.~Rey-Escudero, A.~G. Marques, and D.~Ram\'irez,
  ``Demixing and blind deconvolution of graph-diffused sparse signals,'' in
  \emph{IEEE Int. Conf. on Acoustics, Speech and Signal Process.}, Apr. 2018,
  pp. 4189--4193.

\bibitem{tong_1991_2order}
L.~Tong, G.~Xu, and T.~Kailath, ``A new approach to blind identification and
  equalization of multipath channels,'' in \emph{Asilomar Conf. Signals,
  Systems, and Comp.}, Nov. 1991.

\bibitem{tong_1994_blind}
------, ``Blind identification and equalization based on second-order
  statistics: A time domain approach,'' \emph{IEEE Trans. Info. Theory},
  vol.~40, no.~2, pp. 340--349, Mar. 1994.

\bibitem{Benveniste1980}
A.~Benveniste, M.~Goursat, and G.~Ruget, ``Robust identification of a
  nonminimum phase system: Blind adjustment of a linear equalizer in data
  communications,'' \emph{IEEE Trans. Auto. Control}, vol.~25, no.~3, pp.
  385--399, Jun. 1980.

\bibitem{Ding1991}
Z.~Ding, R.~A. Kennedy, B.~D.~O. Anderson, and C.~R. Johnson, ``Ill-convergence
  of {Godard} blind equalizers in data communication systems,'' \emph{IEEE
  Trans. Commun.}, vol.~39, no.~9, pp. 1313--1327, Sept. 1991.

\bibitem{Tugnait1987}
J.~K. Tugnait, ``Identificaiton of linear stochastic system via second and
  fourth-order cumulant matching,'' \emph{IEEE Trans. Info. Theory}, vol.~33,
  no.~3, pp. 393--407, May 1987.

\bibitem{shalvi_1990_new}
O.~Shalvi and E.~Weinstein, ``New criteria for blind deconvolution of
  nonminimum phase systems (channels),'' \emph{IEEE Trans. Info. Theory},
  vol.~36, no.~2, pp. 312--321, Mar. 1990.

\bibitem{Hatzinakos1989}
D.~Hatzinakos and C.~L. Nikias, ``Estimation of multipath channel response in
  frequency selective channels,'' \emph{IEEE J. Sel. Areas Commun.}, vol.~7,
  no.~1, pp. 12--19, Jan. 1989.

\bibitem{Petropulu1993}
A.~P. Petropulu and C.~L. Nikias, ``Blind convolution using signal
  reconstruction from partial higher order cepstral information,'' \emph{IEEE
  Trans. Signal Process.}, vol.~41, no.~6, pp. 2088--2095, Jun. 1993.

\bibitem{giannakis_1989_identification}
G.~B. {Giannakis} and J.~M. {Mendel}, ``Identification of nonminimum phase
  systems using higher order statistics,'' \emph{IEEE Trans. Acoustics, Speech,
  and Signal Process.}, vol.~37, no.~3, pp. 360--377, March 1989.

\bibitem{giannakis_1989_cumulant}
G.~B. {Giannakis}, Y.~{Inouye}, and J.~M. {Mendel}, ``Cumulant based
  identification of multichannel moving-average models,'' \emph{IEEE Trans.
  Auto. Control}, vol.~34, no.~7, pp. 783--787, July 1989.

\bibitem{liu_1993_deterministic}
H.~Liu, G.~Xu, and L.~Tong, ``A deterministic approach to blind equalization,''
  in \emph{Asilomar Conf. Signals, Systems, and Comp.}, Nov 1993, pp. 751--755.

\bibitem{liu1994}
------, ``A deterministic approach to blind identification of multi-channel
  {FIR} systems,'' in \emph{IEEE Int. Conf. on Acoustics, Speech and Signal
  Process.}, Apr. 1994.

\bibitem{xu_1995_least}
G.~Xu, H.~Liu, L.~Tong, and T.~Kailath, ``A least-squares approach to blind
  channel identification,'' \emph{IEEE Trans. Signal Process.}, vol.~43,
  no.~12, pp. 2982--2993, Dec. 1995.

\bibitem{Moulines1995}
E.~Moulines, P.~Duhamel, J.~Cardoso, and S.~Mayrargue, ``Subspace methods for
  the blind identification of multichannel fir filters,'' \emph{IEEE Trans.
  Signal Process.}, vol.~43, no.~2, pp. 516--525, Feb. 1995.

\bibitem{Slock1994}
D.~T.~M. Slock, ``Blind fractionally-spaced equalization,
  perfect-reconstruction filter banks and multichannel linear prediction,'' in
  \emph{IEEE Int. Conf. on Acoustics, Speech and Signal Process.}, Apr. 1994.

\bibitem{Baccala1994}
L.~A. Baccala and S.~Roy, ``A new blind time-domain channel identification
  method based on cyclostationarity,'' in \emph{Proc. 26th Conf. Informations
  Sciences and Systems}, Mar. 1994.

\bibitem{liu_1996_recent}
H.~Liu, G.~Xu, L.~Tong, and T.~Kailath, ``Recent developments in blind channel
  equalization: From cyclostationarity to subspaces,'' \emph{Signal Process.},
  vol.~50, no. 1-2, pp. 83--99, Apr. 1996.

\bibitem{zhang_2016_one}
H.~Zhang, M.~Yan, and W.~Yin, ``One condition for solution uniqueness and
  robustness of both l1-synthesis and l1-analysis minimizations,'' \emph{Adv.
  in Comp. Math.}, vol.~42, no.~6, pp. 1381--1399, Dec. 2016.

\bibitem{random_graphs}
B.~Bollob\'{a}s, \emph{Random graphs}.\hskip 1em plus 0.5em minus 0.4em\relax
  Cambridge University Press, 2001.

\bibitem{cvx}
M.~Grant and S.~Boyd, ``{CVX}: Matlab software for disciplined convex
  programming, version 2.1,'' \url{http://cvxr.com/cvx}, Mar. 2014.

\bibitem{zachary1977information}
W.~W. Zachary, ``An information flow model for conflict and fission in small
  groups,'' \emph{Journal of anthropological research}, vol.~33, no.~4, pp.
  452--473, 1977.

\end{thebibliography}

\end{document}